\documentclass{aa}

\usepackage{graphicx}
\usepackage{txfonts}
\begin{document} 

   \title{Absolute colors and phase coefficients of asteroids}

   %\subtitle{Small phase angles}

   \author{A. Alvarez-Candal
          \inst{1,2,3}
          \and
          S. Jim\'enez Corral
          \inst{4}
          \and
          M. Colazo
          \inst{5}
          %\and
          %T. Santana-Ros
          %\inst{4,5}%\fnmsep\thanks{Just to show the usage
          %of the elements in the author field}
          }

   \institute{
   Instituto de Astrof\'isica de Andaluc\'ia, CSIC, Apt 3004, E18080 Granada, Spain\\
                 \email{varobes@gmail.com}
    \and
    Instituto de F\'isica Aplicada a las Ciencias y las Tecnolog\'ias, Universidad de Alicante, San Vicent del Raspeig, E03080, Alicante, Spain
    \and
    Observat\'orio Nacional / MCTIC, Rua General Jos\'e Cristino 77, Rio de Janeiro, RJ, 20921-400, Brazil
    \and
    Universidad Internacional de Valencia, Valencia, Spain
    \and
    Instituto de Astronom\'ia Te\'orica y Experimental, CONICET-UNC, Laprida 854, C\'ordoba, Argentina
    %Departamento de F\'isica Ingenier\'ia de Sistemas y Teor\'ia de la Se\~nal, Universidad de Alicante, San Vicent del Raspeig, E03080, Alicante, Spain
    %\and
    %Institut de Ci\`encies del Cosmos (ICCUB), Universitat de Barcelona (IEEC-UB), Mart\'i Franqu\`es 1, E08028, Barcelona, Spain
    %\email{c.ptolemy@hipparch.uheaven.space}
    %\thanks{The university of heaven temporarily does not accept e-mails}
             }

   \date{Received XX; accepted XX}

% \abstract{}{}{}{}{} 
% 5 {} token are mandatory
 
  \abstract
  % context heading (optional)
  % {} leave it empty if necessary  
  {We use phase curves of small bodies to measure absolute magnitudes and, together with complementary theoretical and laboratory results, to understand their surfaces' micro and macroscopic properties. Although we can observe asteroids up to phase angles of about 30 deg, the range of phase angles covered by outer solar system objects usually does not go further than 7 to 10 deg for centaurs and 2 deg for trans-Neptunian objects, and a linear relation between magnitude and phase angle may be assumed.}
  % aims heading (mandatory)
  {We aim at directly comparing data taken for objects in the inner solar system (inside the orbit of Jupiter) with data of centaurs and trans-Neptunian objects.}
  % methods heading (mandatory)
  {We use the SLOAN Moving Objects Catalog data to construct phase curves restricted to phase angles less than or equal to 7.5 deg, compatible with the angles observed for the trans-Neptunian/Centaur population. We assume a linear model for the photometric behavior to obtain absolute magnitudes and phase coefficients in the ugirz, V, and R filters.}
  % results heading (mandatory)
  {We obtained absolute magnitudes in seven filters for $>4000$ objects. Our comparison with outer solar system objects points to a common property of the surfaces: intrinsically redder objects become blue with increasing phase angle, while the opposite happens for intrinsically bluer objects. }
  % conclusions heading (optional), leave it empty if necessary 
  {}

   \titlerunning{Absolute colors and phase coefficients of asteroids}
   \authorrunning{Alvarez-Candal et al.}

   \maketitle
%
%-------------------------------------------------------------------

\section{Introduction}\label{sect:intro}
Small bodies show variations in their brightness. These variations are periodic and usually approximate a series of sines and cosines \cite[see for example][]{harrislupishkp1989aste}. Some of these variations are related to the body's rotation along its spin axis. Other brightness variations are related to the changing geometry of the system formed by the Sun, the small body, and the Earth. We are interested in this last kind of brightness variation.

Geometrically speaking, brightness variations happen because of the changing distances of the object to the Sun and the Earth and the fraction of surface illuminated as seen from the Earth. If we disregard for the time being the rotational variations, we can correct the effect of the distances using the reduced magnitudes:
\begin{equation}\label{eq:0}
    M(\alpha)=M-5log{(D\Delta)},
\end{equation}
where $M$ is the observed apparent magnitude, in any filter, $D$ is the distance object-Sun, and $\Delta$ is the distance object-Earth (both in astronomical units). The phase angle $\alpha$ is the angular distance between the Earth and the Sun as seen from the small body. Once we remove the distances' effect, we can study the brightness change due to the changing $\alpha$ using phase curves.

For the scope of this work, two previous studies are of particular interest, and we will devote a few lines to each of them. In \cite{alcan2019}, we presented a catalog of phase curves in two filters, V and R, {of 117 trans-Neptunian objects and centaurs.} There, we discovered a strong anti-correlation between the absolute colors and the relative phase coefficients (we will define both quantities in Sect. \ref{sect:analysis}), indicating that redder objects have steeper phase curves in the R filter than in the V filter, or, in other words, redder objects tend to become bluer with increasing $\alpha$. This work showed that phase curves in more than one wavelength are essential tools usually overlooked (the only exception we know of being \citealt{mahlke2021}). On the other hand, in \citet[hereafter AC22]{alcan2022}, we obtained multi-wavelength phase curves for asteroids using the data from the Moving Objects Catalog (MOC) of the SLOAN Digital Sky Survey for over 14\,000 asteroids. We used the HG$_{12}^*$ model as implemented by \cite{penti2016HG}, obtaining the probability distributions of the absolute magnitudes and phase coefficients for each filter.

{We used the same input data as in AC22 to test whether a similar anti-correlation between absolute colors and phase coefficients exists for asteroids. We use the same selection criteria as in \cite{alcan2019}: at least three data and $\alpha\in[0,7.5]$ deg; the range in $\alpha$ covers that spanned by centaurs ($<7$ or 8 deg) and TNOs $<2$ deg. We assumed a simple linear model to make a fair comparison with \cite{alcan2019}. Note that this range of $\alpha$ is slightly lower than that covered by their data. We stress that in the region $\alpha\leq7.5$ deg, we are within the Opposition Effect regime due to a combination of shadow-hiding \citep{hapke1963JGR} and coherent back-scattering \citep{muino1989OE}. This effect is characterized by an apparent surge in brightness starting at about $\alpha=5$ to 10 deg and extending to its maximum at opposition, i.e., $\alpha\approx0$ deg. {Note, however, that in some TNOs, it may start as low as $\alpha\approx0.1$ deg \citep{verbi2022}. In Appendix \ref{appA}, we will tackle the validity of the linear approach for our work.}

{We organized this work as follows:} In Sect. \ref{sect:data} we briefly present the dataset we used. In Sect. \ref{sect:analysis} we describe the procedure we applied, while in Sects. \ref{sect:discussion} and \ref{sect:conclussions} the results are discussed and the conclusions of this work are summarized.}

\section{Dataset}\label{sect:data}
\begin{figure}
\centering
\includegraphics[width=\hsize]{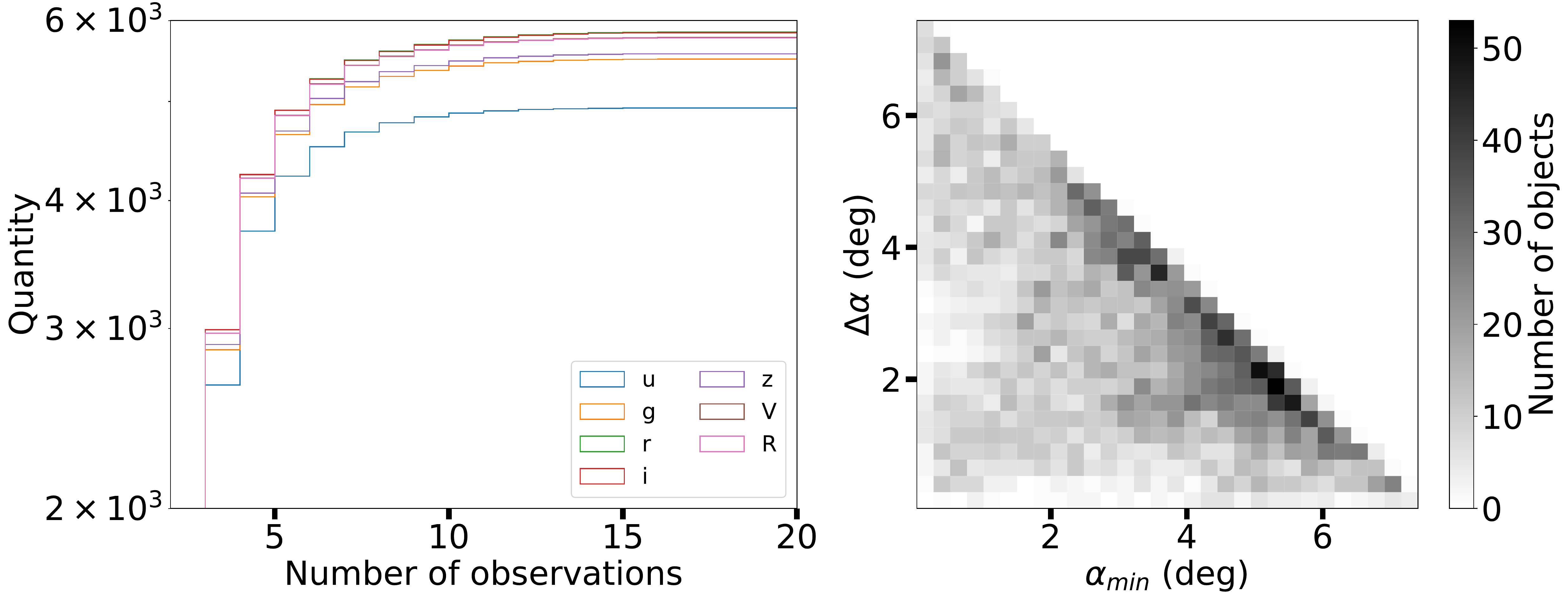}
\caption{Observational circumstances of the data used in this work. Left panel: Cumulative distribution of the number of observations per filter. Right panel: Minimum $\alpha$ vs. span in $\alpha$.}\label{fig:fig02}%
\end{figure}

We use the SLOAN MOC extended with the SVOMOC \citep[see][]{ivezic2001AJ,juric2002AJ,carry2016Icar}. We will call this MOC for simplicity. In a nutshell, the database contains 277\,747 $u'$, $g'$, $r'$, $i'$, $z'$ magnitudes, and their errors\footnote{We note that we drop the $'$ in the remaining of the text.}, of 141\,388 moving objects. In the rest of the text, we use $m$ to refer to the apparent magnitude in any SLOAN filters, the V or R filters, unless explicitly stated. When naming a filter, we use italics when speaking about magnitudes and regular text. Our selection criteria were: (i) data with $\alpha\leq7.5$ deg, (ii) at least three data, and (iii) $m : \sigma_{m}\leq1$. These criteria produced a total of 5\,848 objects, most of these with $\alpha\in(2.5,5.5)$ deg, but there are a few objects that cover the whole interval, while most objects have only three observations (Fig. \ref{fig:fig02}). 

\section{Analysis}\label{sect:analysis}
Phase curves are exciting and potent tools for obtaining different information, especially absolute magnitudes ($H$) and phase coefficients. The absolute magnitude relates to the amount of light reflected by the object with its reflective area (illuminated) and how the light is reflected (albedo). On the other hand, the phase coefficients describe the shape of the curve and tell about the scattering properties of the surface, although in no clear nor direct {manner \citep[see, for instance,][]{lummebowell1981AJ,hapke2002}}. The phase coefficients and their physical interpretation depend on the adopted photometric model, which depends on the data quality and how well-sampled the phase curve is. The model adopted by the International Astronomical Union is the HG$_1$G$_2$ \citep{muinonen2010HG1G2}, with improvements made to work with low-quality measurements and sparsely covered curves in the HG$_{12}^*$ model \citep{penti2016HG}. In our case, we will use a simpler two-parameter model, assuming that the phase curves are linear:
\begin{equation}\label{eq:1}
    M(\alpha) =  H+\beta \alpha,
\end{equation}
where $M(\alpha)$ is the reduced magnitude shown in Eq. \ref{eq:0}, $H$ is the absolute magnitude, and $\beta$ is the phase coefficient in units of mag deg$^{-1}$. We use the linear model for all the filter sets considered in this work {(see Appendix \ref{appA}).}

\begin{figure}
\centering
\includegraphics[width=\hsize]{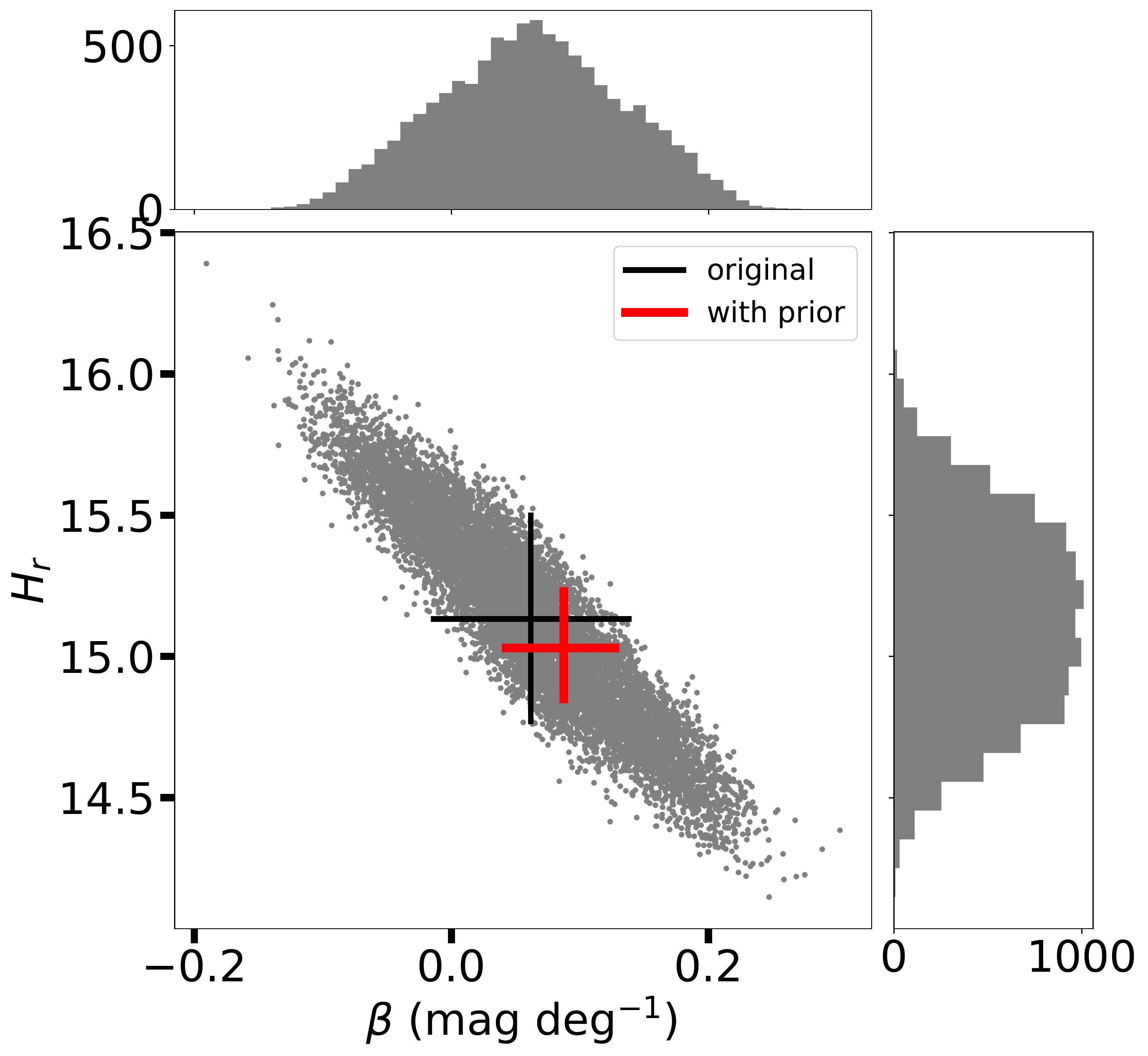}
\caption{Example of the density distribution obtained for asteroid 1034 T-1. The black cross shows the adopted $H_r$ and $\beta_r$ before applying the prior in magnitudes, while the red cross shows the final values.}\label{fig:fig01}%
\end{figure}

We follow the method developed in AC22, using their probability distributions of possible rotational states  $P(\Delta m|H_V)$ (see their Eq. 2) for the objects in common while computing our own whenever necessary. Then we perform a Monte Carlo simulation to estimate the absolute magnitudes. The nominal values are the median of the distributions, while the uncertainty interval is between the 16$^{\rm th}$ and 84$^{\rm th}$ percentile (see Fig. \ref{fig:fig01}).
The figure shows that the distributions in $H_r$ and $\beta_r$ are broad, producing large uncertainties (seen as the black cross in the plot). In total, we have results for more than 4500 objects in all five filters of the MOC and also in the V and R filters (Table \ref{table:1}). These last two are obtained transforming the ugriz magnitudes into Johnson-Cousins magnitudes \citep[see Table 1 in][for stars with $R - I < 1.15$]{jester2005AJ}\footnote{ http://www.sdss3.org/dr8/algorithms/sdssUBVRITransform.php}.
\begin{table}
\caption{Number of absolute magnitudes obtained}
\label{table:1}
\centering
\begin{tabular}{c c | c c}
\hline\hline
Concept  & N         & Concept & N \\
\hline
$H_u$   & 4\,924   & $H_V$                  & 5\,770 \\ 
$H_g$   & 5\,496   & $H_R$                  & 5\,760 \\
$H_r$   & 5\,840   & at least one valid $H$ & 5\,848 \\
$H_i$   & 5\,831   & all five valid $H$     & 4\,529 \\
$H_z$   & 5\,561   & $H_V-H_R$              & 5\,759 \\
\hline
\end{tabular}
\end{table}

\subsection{Using AC22's results as prior}
\begin{figure}
\centering
 \includegraphics[width=4.0cm]{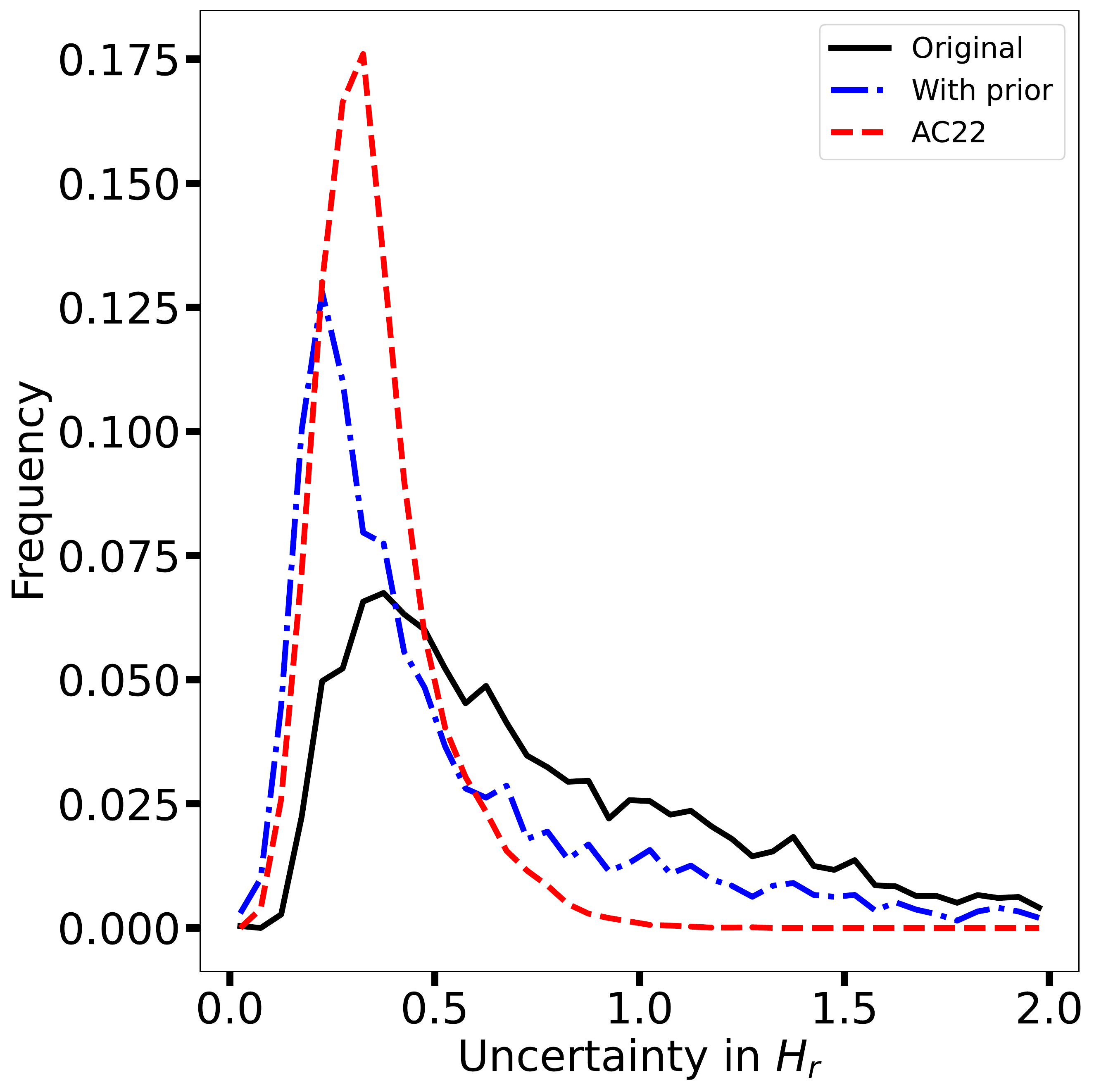}
 \includegraphics[width=4.0cm]{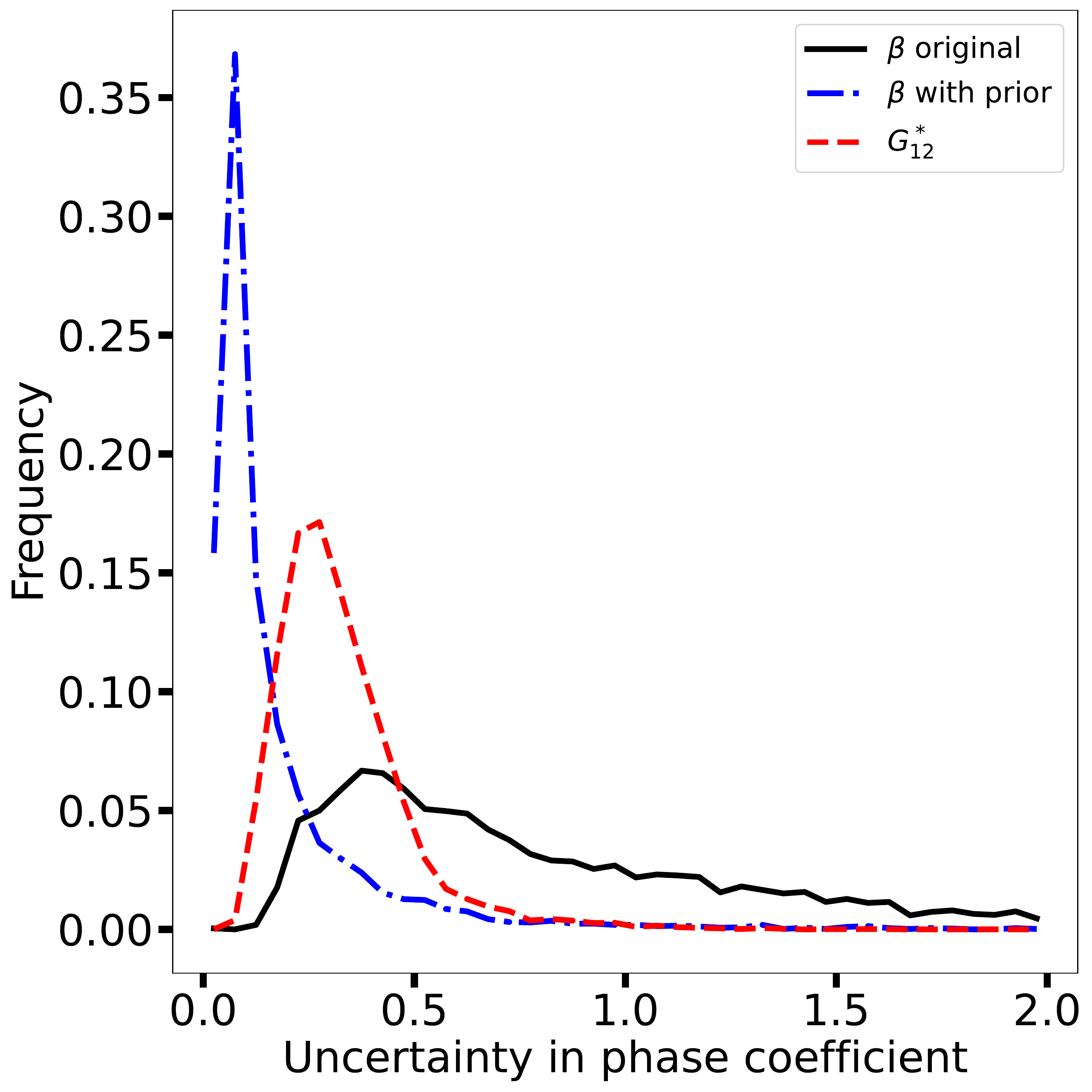}
\caption{Distributions of uncertainties. The red dashed line indicates AC22. The continuous black line shows the uncertainties before applying AC22's prior, and the blue dot-dashes line the final uncertainty distribution after applying the prior. In the left panel are shown the uncertainties in $H_r$, while the right panel shows them for the phase coefficient. Note that in the latter, the coefficients differ between AC22 and ours.}\label{fig:fig03}%
\end{figure}

As hinted in Fig. \ref{fig:fig01}, the uncertainties in $H$ and $\beta$ tend to be relatively large, especially when a low number of points covers the phase curve (see the impact in Appendix A of \citealt{alcan2019}). Nevertheless, it is possible to improve them, at least for a significant number of objects. In AC22, we obtained solutions of $H$ for over 14\,000 objects, in the form of probability distributions, with a different photometric model. We use these probability distributions as priors to improve our results. We used the probabilities in $H_{AC22}$, $P(H_{AC22})$, in two steps:
\paragraph{Absolute magnitudes:} In this case, the re-assignation of probabilities was performed directly in the absolute magnitudes space:
\begin{equation}\label{eq:2}
    P_i(H)=\frac{{P_i(H_{original}) P_i(H_{AC22})}}{\sum_j{P_j(H_{original}) P_j(H_{AC22})}},
\end{equation}
where $P(H_{original})$ is the probability distribution obtained in this work, for example, the right histogram in Fig. \ref{fig:fig01}. We use a bin width of 0.02 magnitudes in the interval from 8 to 24 for the probability distributions. We ran the process for 3206 objects.
\paragraph{Phase coefficients:} In AC22, the phase coefficients were the $G_{12}^*$, which are not comparable to $\beta$. Therefore their $P(G_{12}^*)$ cannot be used as priors here. In this case, we will use a similar approach as in AC22 and use the $P(\beta|H_{original})$ as follows:
\begin{equation}\label{eq:3}
    P_i(\beta)=\frac{\sum_j{P_i(\beta|P^j(H_{original}))P^j(H_{AC22})}}
    {\sum_i{\sum_j{P_i(\beta|P^j(H_{original}))P^j(H_{AC22})}}},
\end{equation}
where $P(\beta|P^j(H_{original}))$ is binned between -3 and 3 mag deg$^{-1}$ with a bin width of 0.005 mag deg$^{-1}$. The application of the priors from AC22 led to smaller uncertainties and slightly different nominal values (see Figs. \ref{fig:fig01},  \ref{fig:fig03}, and Table \ref{table:2}).
\begin{table}
\caption{Median uncertainties}
\label{table:2}
\centering
\begin{tabular}{c c c }
\hline\hline
Reference   & in $H_r$ & in Phase coefficient      \\
\hline
AC22                     & 0.33 & 0.30\tablefootmark{a}    \\
This Work (original)     & 0.70 & 0.71 (mag deg$^{-1}$)\\
This Work (w/prior)      & 0.39 & 0.09 (mag deg$^{-1}$)\\
\hline
\end{tabular}
\tablefoot{The first column indicates the reference, the second column the median value of the uncertainty in $H_r$, and the third column shows the median value of the slope parameter.
{\tablefoottext{a}{Corresponds to $G_{12}^*$ and does not have units.}}}
\end{table}
{From this point on, our database is composed of the updated magnitudes for the 3206 objects with priors available plus 2642 objects which did not, to keep the total of 5848 objects.}

\subsection{Colors and $\Delta\beta$}\label{sect:colors}
An essential part of this work is the comparison of colors and $\Delta\beta$ (also called relative phase coefficients in other parts of the text). Given two quantities represented by their probability distributions $P_1$ and $P_2$, the nominal value of the difference is the difference in the nominal values. On the other hand, the uncertainty range is not determined by the expected propagation of errors as it does not consider the complete information contained in the distributions. In this case, we opted to compute the probability distribution of the difference between two quantities $X$ and $Y$ as $Z=X-Y$ 
\begin{equation}\label{eq:4}
    \implies P(Z)=\big(P(X)*P(Y)\big).
\end{equation}
Therefore, in the case of a color $C_{ij}=M_i-M_j$ or $\Delta\beta_{ij}=\beta_i-\beta_j$, we obtain the probability distributions $P(C_{ij})$ and $P(\Delta\beta_{ij})$ where the uncertainty intervals are determined, as before, as the 16$^{\rm th}$ and 84$^{\rm th}$ percentiles.

\begin{figure}
\centering
\includegraphics[width=\hsize]{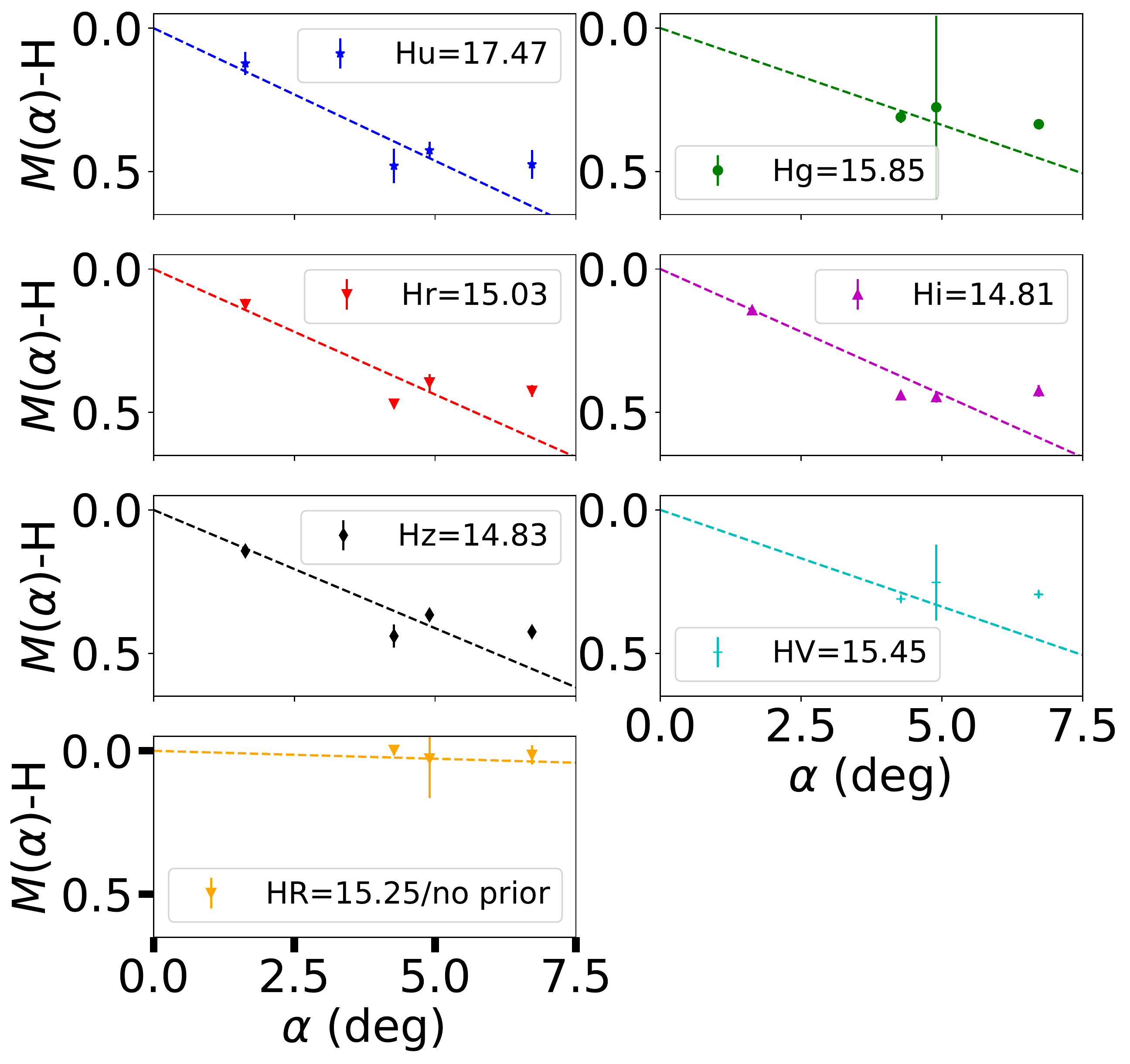}
\caption{Phase curves of asteroid 1034 T-1. Different panels show data in different filters (labeled in each inset).}\label{fig:example}%
\end{figure}

The final results of $H$ and $\beta$ are shown in Table \ref{table:3} below. {As noted above, these are values after applying AC22 priors. We also draw attention to the fact that the table shows median values of a probability distribution, which may not be the optimal solution (for example, one obtained by applying a minimization algorithm).}
\begin{table*}
\caption{Sample of the results}
\label{table:3}
\centering
\begin{tabular}{c c c c c c c c c c}
\hline\hline
ID   & $H_u$ & $\sigma^-_{H_u}$ & $\sigma^+_{H_u}$ & N & $\beta_u$ (mag deg$^{-1}$) & $\sigma^-_{\beta_u}$ (mag deg$^{-1}$) & $\sigma^-_{\beta_u}$ (mag deg$^{-1}$) & $\alpha_{min}$ (deg) & $\Delta\alpha$ (deg) \\
\hline
1034\_T-1 & 17.4700& 	0.1800& 	0.2000& 	4& 	0.0925	 &0.0500&	0.0400 &  1.63 & 5.09 \\
1054\_T-3 & 19.2714& 	0.6733& 	0.6518& 	3& 	0.0126	 &0.2716&	0.2760 &  1.31 & 1.89 \\
1141\_T-2 & 19.2879& 	0.8799& 	0.8949& 	3& 	0.0242	 &0.2762&	0.2718 &  2.13 & 1.78 \\
1162\_T-1 & 16.9100& 	0.2000& 	0.1800& 	3&  -0.0625	 &0.0800&	0.0800 &  1.00 & 3.47 \\
1227\_T-1 & 18.7027& 	0.8866& 	0.8635& 	3& 	-0.0288	 &0.4022&	0.4118 &  0.45 & 2.81 \\
\hline
\end{tabular}
\tablefoot{The first column indicates the object's ID, the second the median value of $H_u$, while the third and fourth show the uncertainties. The fifth column indicates the number of points in the phase curve. The sixth, seventh, and eighth columns show the median value of $\beta_u$ and its uncertainties. The last two columns show the minimum phase angle and the total span in $\alpha$. The full table is available at the CDS or upon request. }
\end{table*}
We show an example of the phase curves in Fig. \ref{fig:example} for asteroid 1034 T-1\footnote{This object was the example used in AC22; thus, we selected it as our example.}
The full table with the results is available at the CDS and upon request. The probability distributions and figures are available upon request.

\section{Results and discussion}\label{sect:discussion}
\begin{figure}
\centering
 \includegraphics[width=4cm]{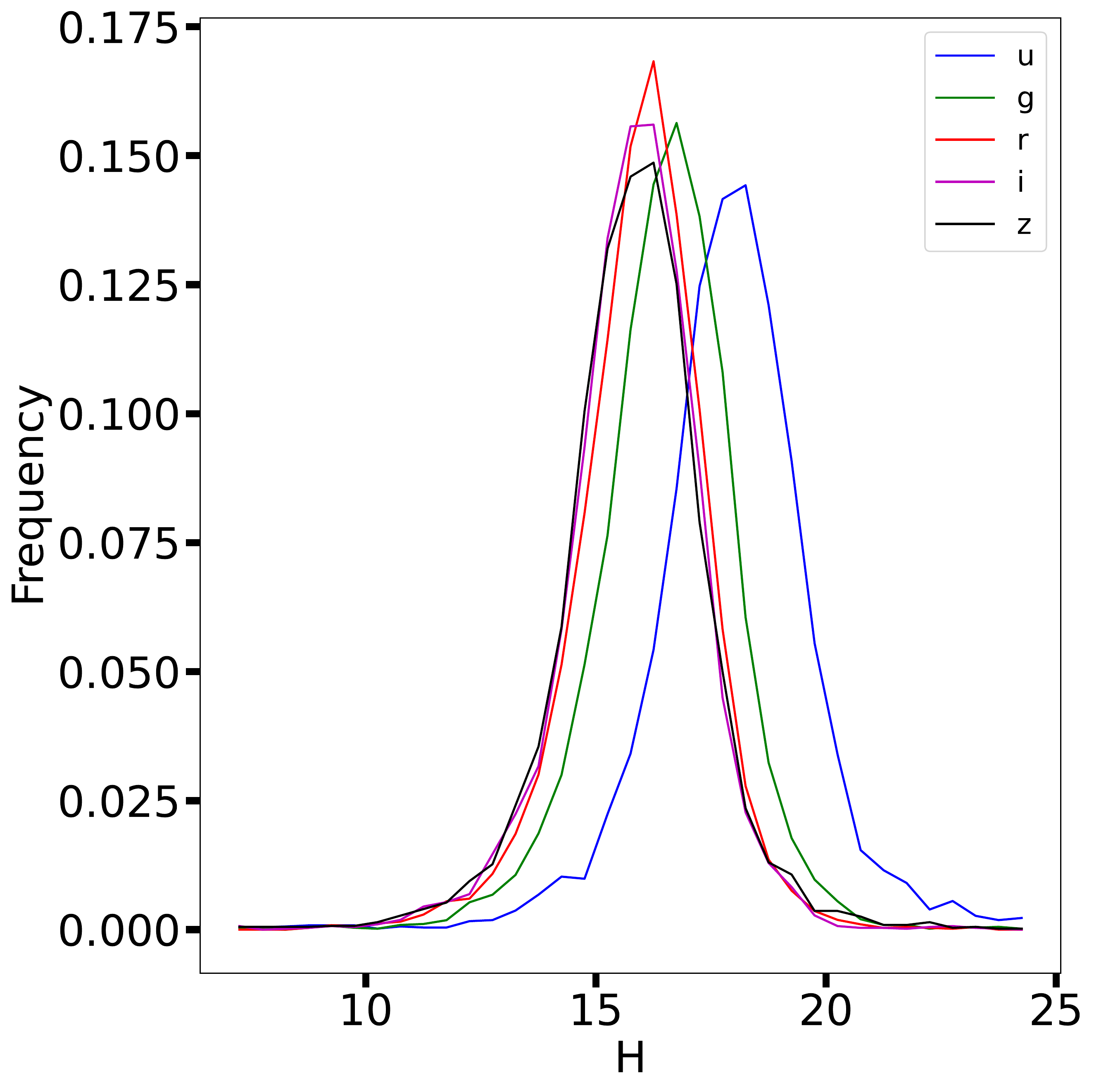}
 \includegraphics[width=4cm]{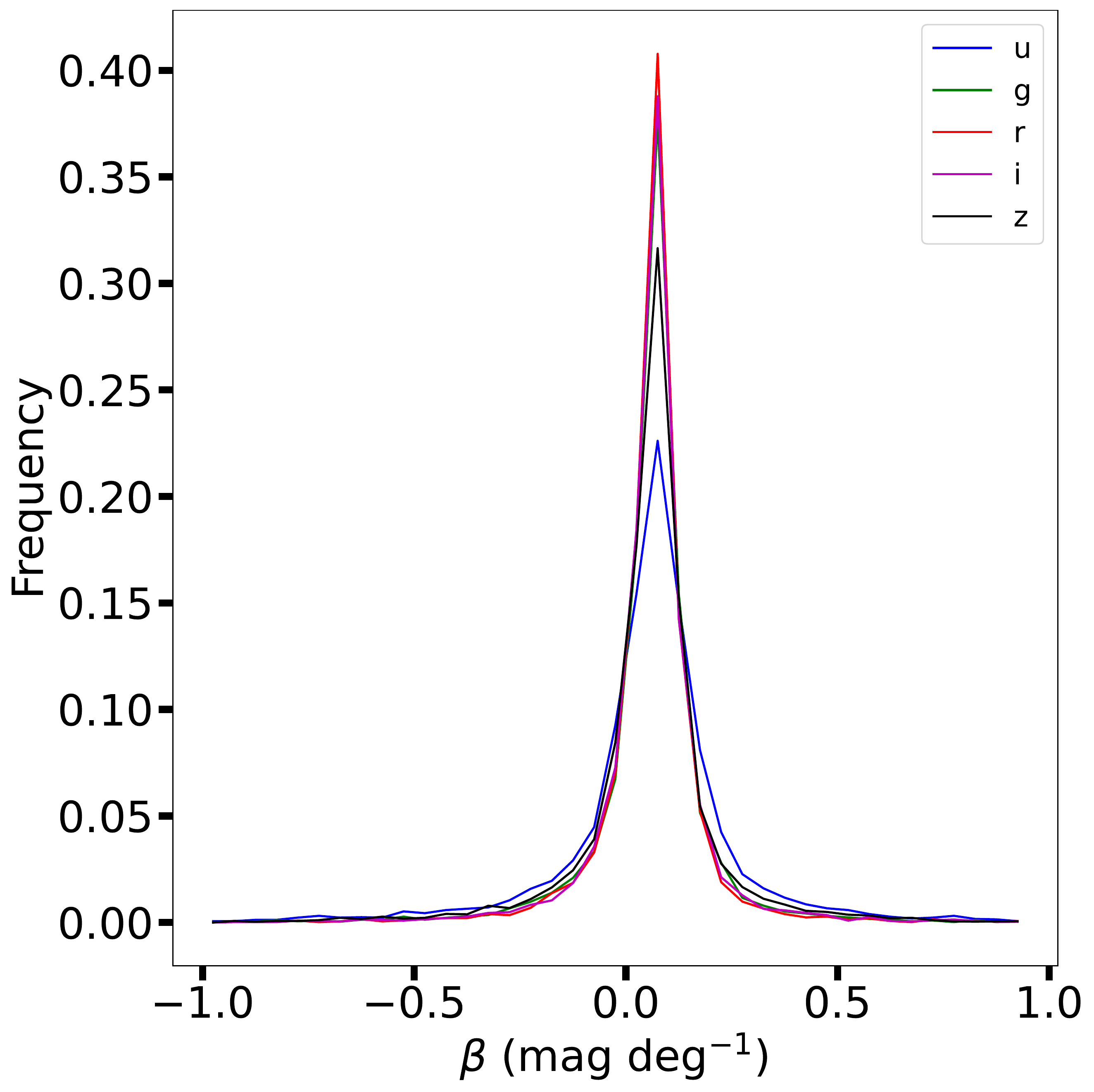}
\caption{Distributions of $H$ (left panel) and $\beta$ (right panel). $u$ is shown in the blue line, $g$ in green, $r$ in red, $i$ in purple, and $z$ in black.}\label{fig:fig07}%
\end{figure}

We show the distributions of $H$ and $\beta$ in the two panels of Fig. \ref{fig:fig07}. In the case of $H$, all filters show similar distributions, except for $H_u$ that peaks at fainter magnitudes than the rest, as discussed in AC22, partly showing the solar $u-g=1.43$; while for $\beta$, the distributions seem centered at similar value (median values fall within 0.068 and 0.071 mag deg$^{-1}$) with standard deviations ranging from 0.13 mag deg$^{-1}$, for $\beta_r$, up to 0.21 mag deg$^{-1}$ for $\beta_u$. These figures are a proxy of the depth of the survey if we only considered objects observed with $\alpha\leq7.5$ degrees.

\begin{figure}
\centering
 \includegraphics[width=4cm]{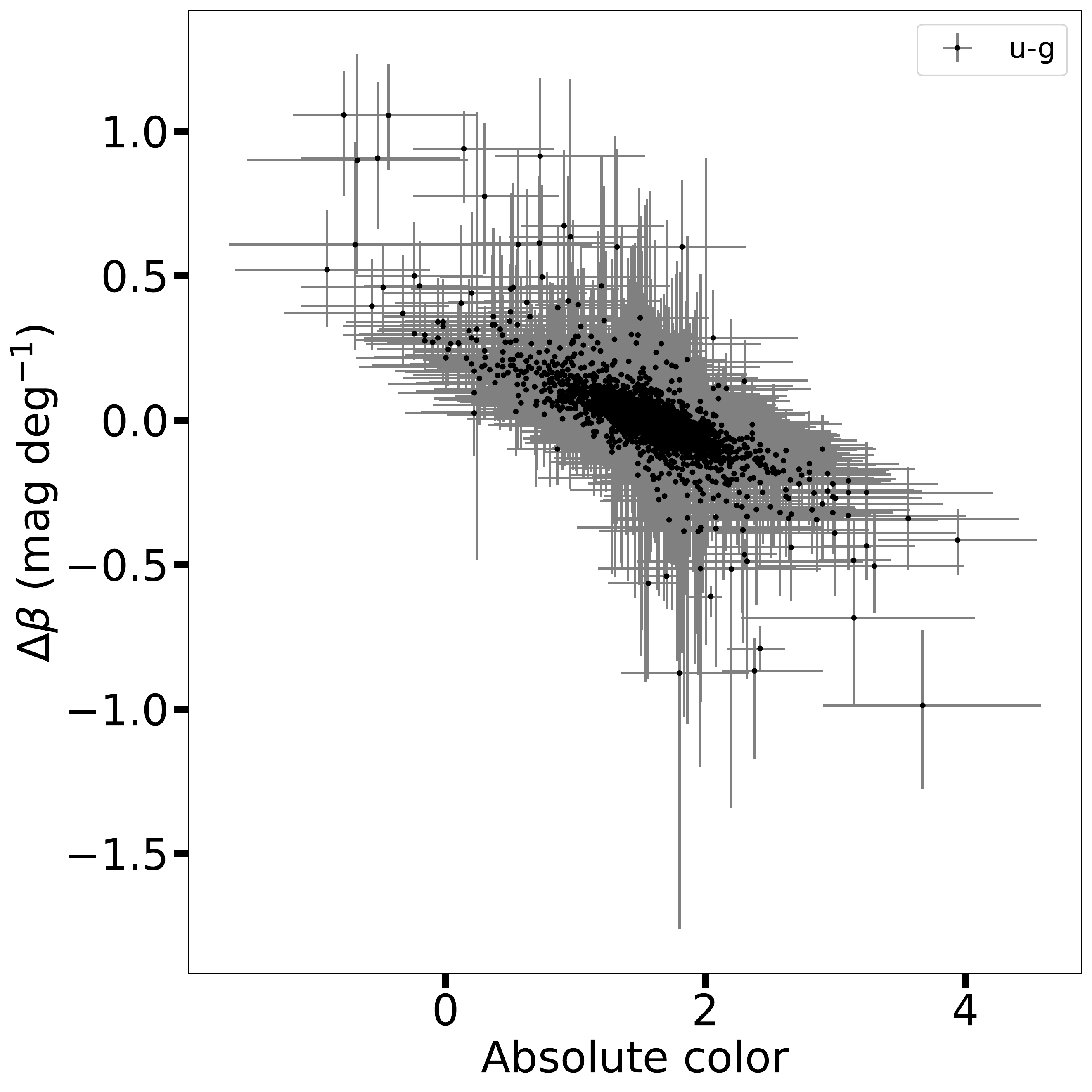}
 \includegraphics[width=4cm]{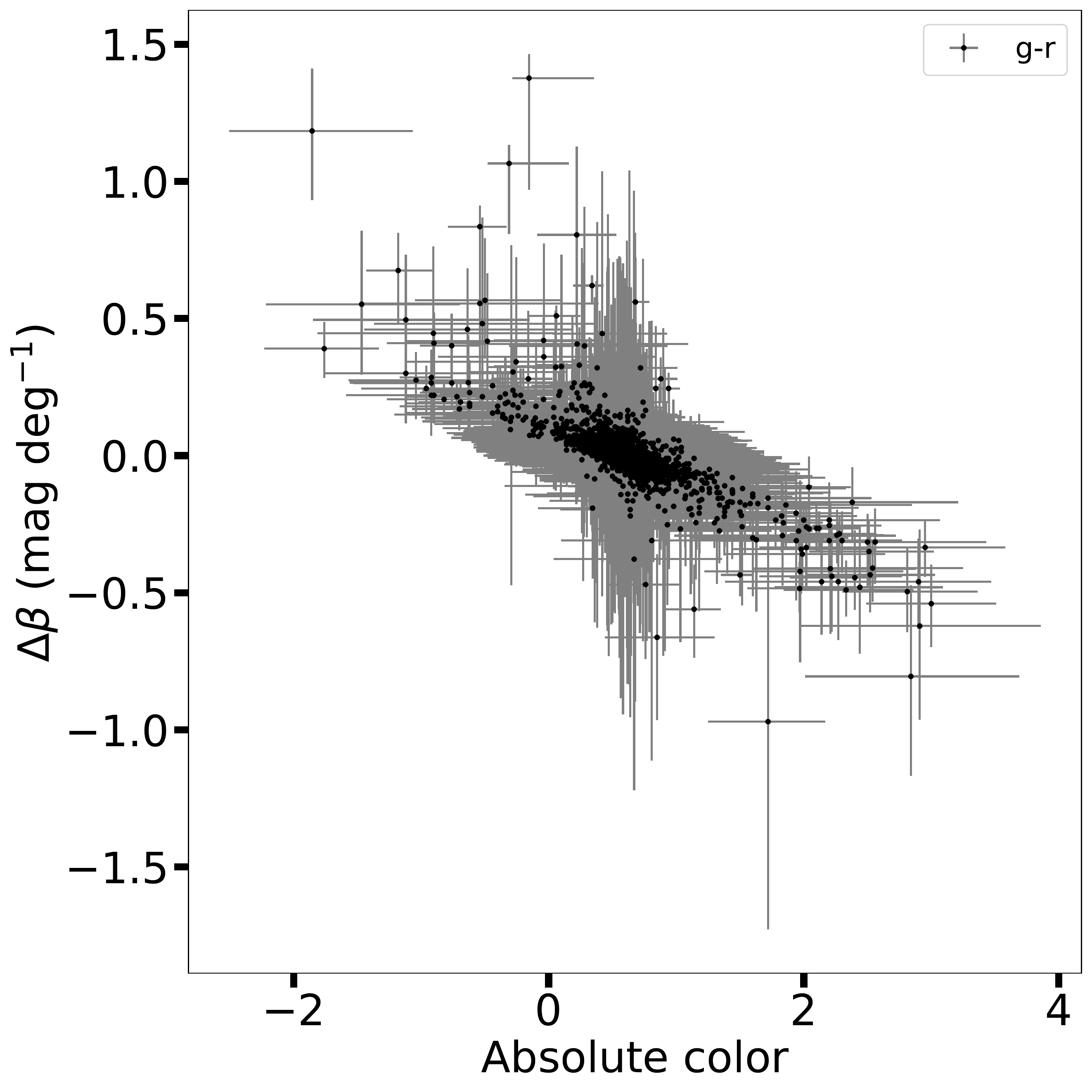}
 \includegraphics[width=4cm]{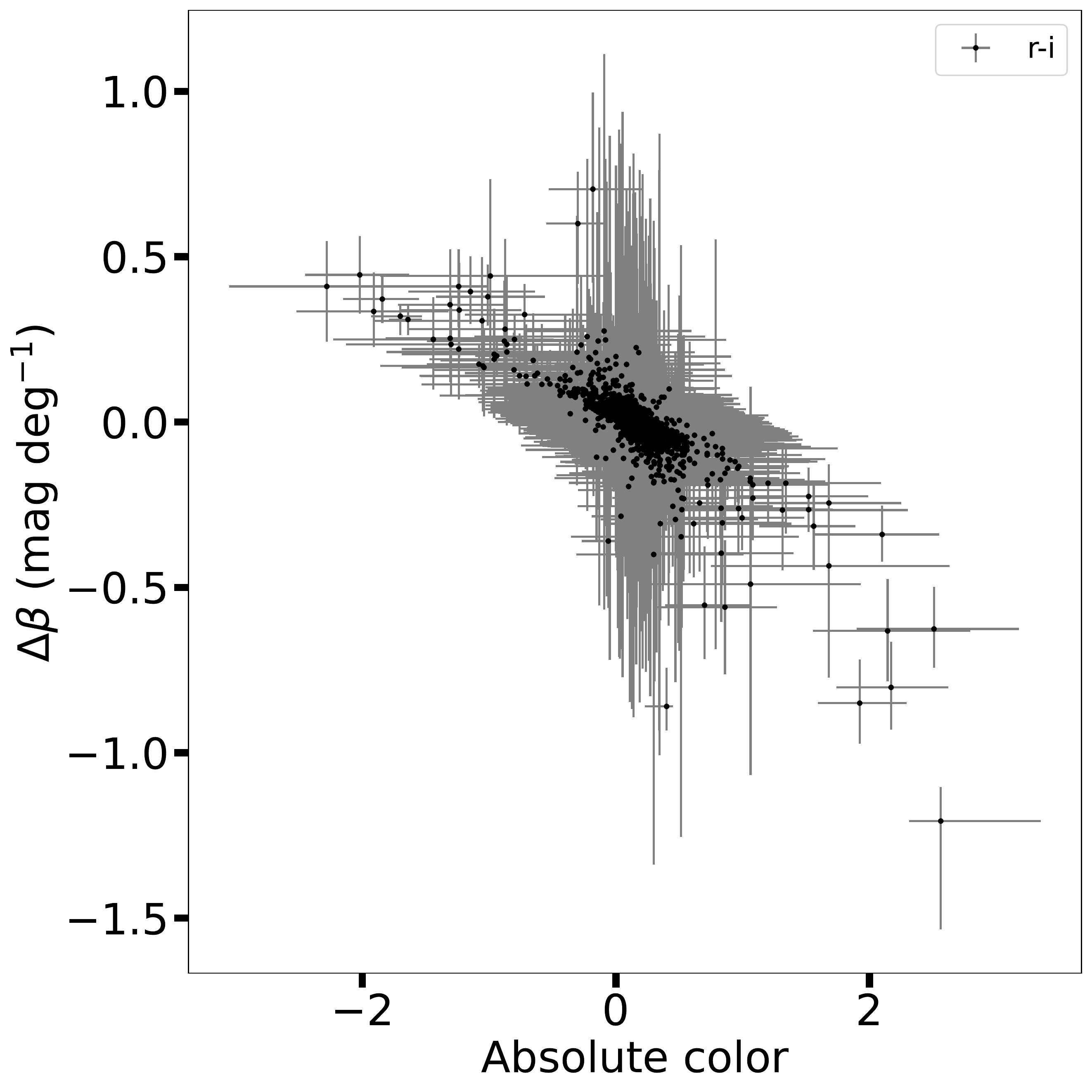}
 \includegraphics[width=4cm]{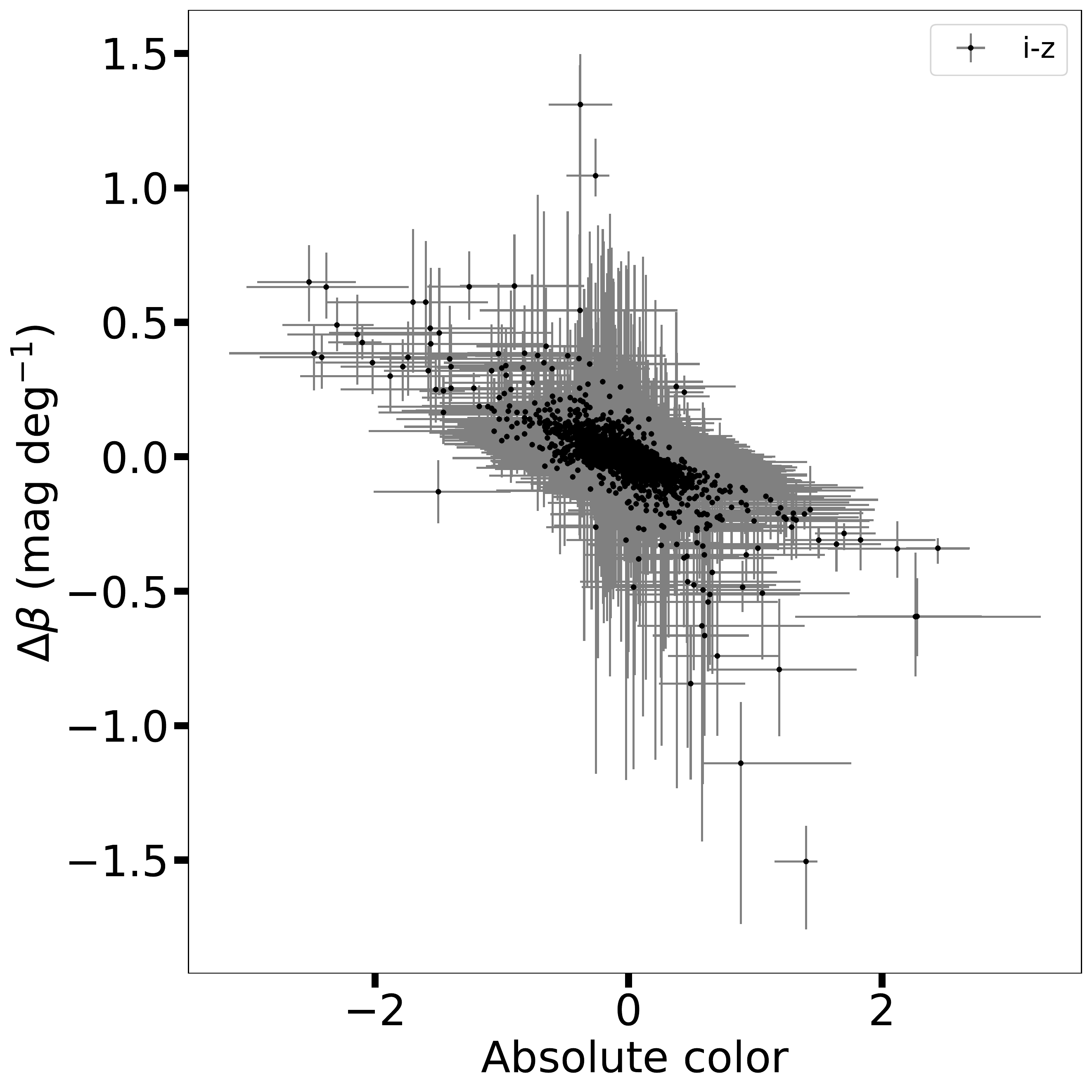}
\caption{Absolute colors versus $\Delta\beta$ for pairs of filters: Top left, $u-g$; top right, $g-r$; bottom left, $r-i$; bottom right, $i-z$.}\label{fig:fig05}%
\end{figure}

The next step is to compute the different filters' absolute colors from $H$ and $\Delta\beta$. Following the procedure outlined above (Sect. \ref{sect:colors}) to assign uncertainties, we created a series of plots shown in Fig. \ref{fig:fig05} for the MOC filters. The figure displays only colors within 2.5 magnitudes of the median of the distribution and with one-sided uncertainty of less than one magnitude.
Following these criteria the figures show 57\% of data in the $u-g$ plot (69\% in $g-r$, 74\% in $r-i$, and 69\% in $i-z$). Note that we did this for clarity purposes only. The bulk of the data follows the main trends discussed below independently of the size of their uncertainties.

All panels in Fig. \ref{fig:fig05} show the same trend: a strong anti-correlation between absolute color and $\Delta\beta$. We run a Spearman test obtaining the rejection of the null hypothesis (that the two variables do not correlate) with high confidence in all cases. The relation indicates that redder objects show steeper phase curves in any pair of filters in the redder filter. Physically, this means that objects with redder colors tend to become bluer with increasing phase angles, while bluer objects tend to get redder with increasing phase angles.

\begin{figure}
\centering
\includegraphics[width=\hsize]{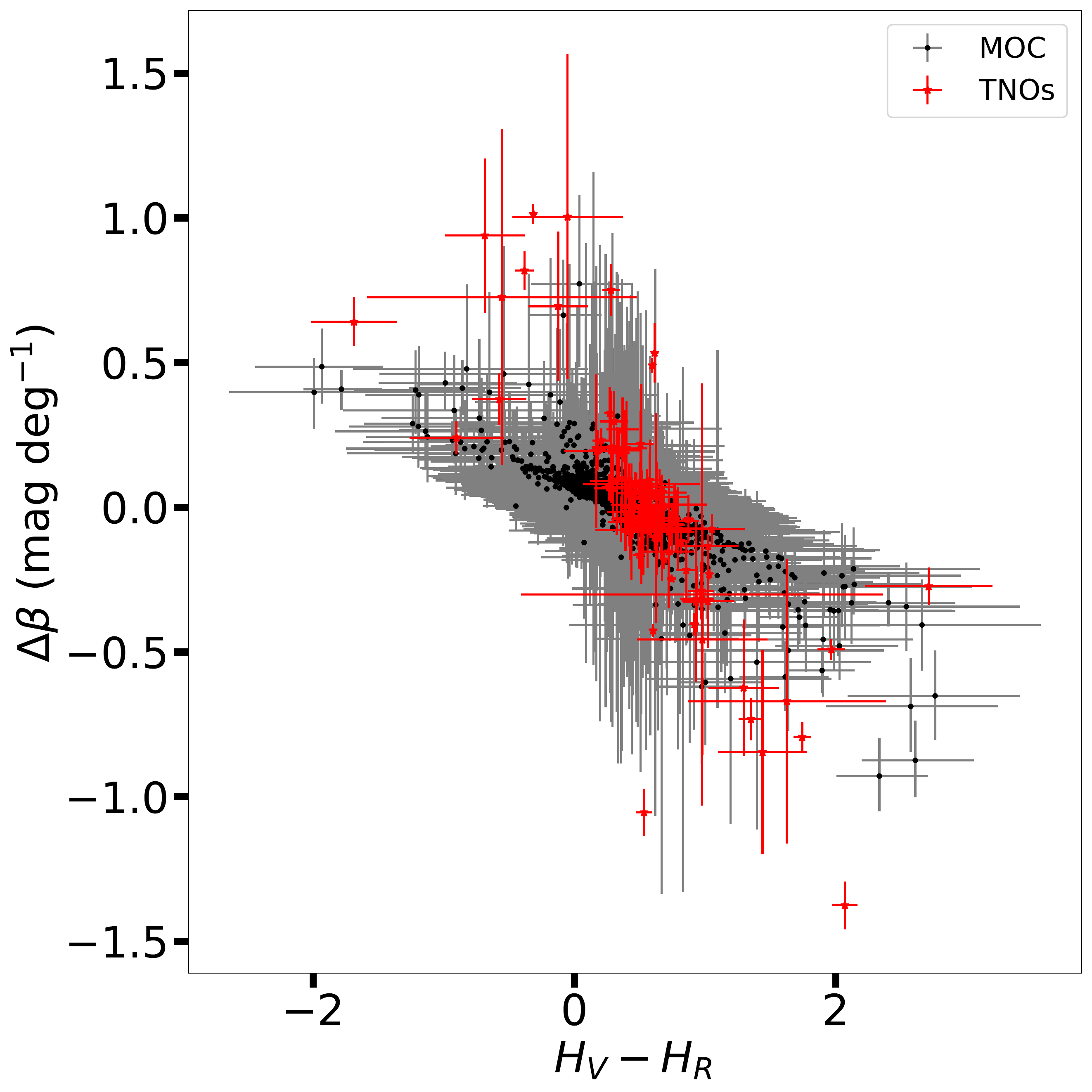}
\caption{$H_V-H_R$ versus $\Delta\beta$. In black dots are shown our results, while in red asterisks are shown the results for TNOs from \cite{alcan2019}.}\label{fig:fig06}%
\end{figure}

This phenomenon was already seen in TNOs by \cite{alcan2019} using $V-R$. To directly compare our results with theirs, we also computed the phase curves in V and R filters transforming from ugriz to UVBRI. The results are in Fig. \ref{fig:fig06} where the data from \cite{alcan2019} is overplotted in red asterisks.
We proposed that this phenomenon may be related to large-sized particles on the surface of (perhaps) icy bodies and that it may show a predominance of single-scattering at a low-phase angle rather than multiple scattering, but probably not related to the surface composition. In the case presented in this work, we are speaking of smaller objects whose surfaces should be volatile-poor, strengthening our hypothesis of no relation with surface composition.

\begin{figure}
\centering
\includegraphics[width=\hsize]{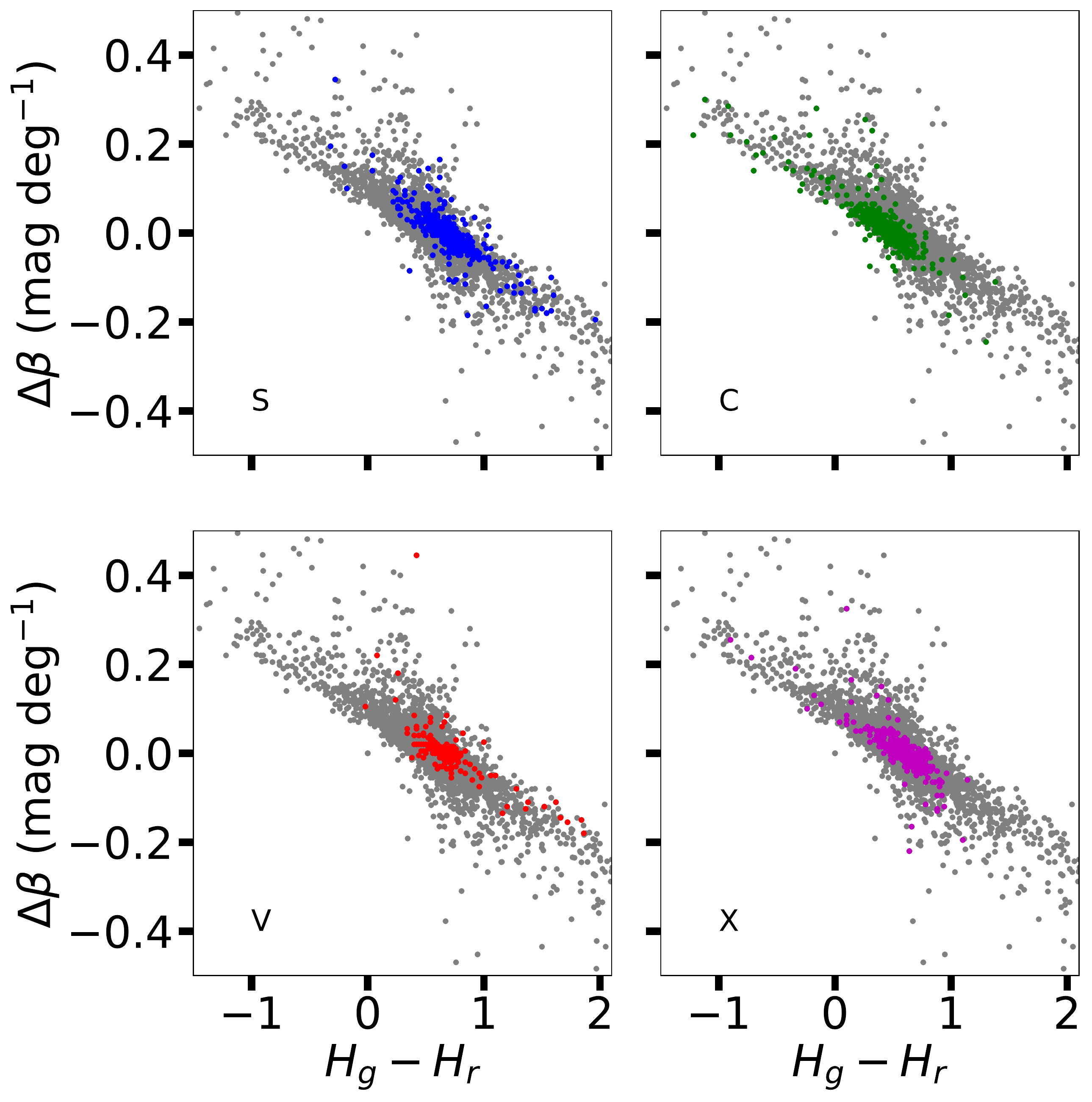}
\caption{$H_g-H_r$ versus $\Delta\beta$. In all panels, the complete sample is shown in gray dots, while different colors mark the different complexes: S-complex in blue (upper left panel), C-complex in green (upper right panel), V-complex in red (bottom left panel), and X-complex in purple (bottom right panel). We do not show the error bars for clarity.}\label{fig:fig08}%
\end{figure}

We also used the taxonomy classification presented by \cite{colazo2022}. In brief, the authors used AC22's $H$ and unsupervised machine learning algorithms to find four clusters in the $H_g-H_i$ vs. $H_i-H_z$ space. They associate each cluster with the four major complexes: The S-complex, the C-complex, the X-complex, and the V-complex. We show the results in Fig. \ref{fig:fig08}, where it is possible to see that all taxa follow the same trend of anti-correlation found for the complete sample. Nevertheless, it is possible to see that each taxon does not distribute similarly. Especially the C- and S-complexes seem to follow parallel trends, with the C-complex slightly on the blue side. The V-complex appears very similar to the S-complex, while the X-complex seems to fall between the C- and S-complex. 
These slight differences are related to each complex characteristic spectra: the C- and X-complex are more linear with no strong absorption features, while the S- and V-complex show adsorptions starting at about 750 nm and redder slopes before the onset of the band. {These results confirm that the relation between absolute colors and relative phase coefficients does not depend on any particular surface composition.}
\smallskip

Could the correlation be possible because of some systematic we are not considering? \cite{beck2021} presented a thorough study of the phase curves of meteoritical material, targeted especially to dark-like material. Their work also detected that objects initially red at a low-phase angle tend to become blue with increasing $\alpha$. However, they reported that the relationship happened on normalized reflectance and that it happened because the shadow hiding effect is (almost) an additive effect and, therefore, the blueing should not be a physical effect. Note that using colors is just a way of showing normalized reflectance. To compare our results to theirs, we proceed as follows: first, we matched our data with the AKARI database \citep{alilagoa2018} and found 59 objects in common. Second, we computed colors relative to $H_r$ for all objects and removed the solar colors to compute relative reflectance using
\begin{equation}
    S^r(\alpha)=\{S_j\} = \{10^{-0.4[(C_{jr}(\alpha)-(j-r)_{\odot}]}\},
\end{equation}
where ${\odot}$ indicates solar colors and
\begin{equation}
    C_{jr}(\alpha)=(H_j-H_r)+\alpha\Delta\beta_{jr}.
\end{equation} 
These values provide relative reflectance normalized to 1 at the effective wavelength of the r filter at a given $\alpha$\footnote{This approximation is only valid for $\alpha\leq7.5$ deg.}. Third, we re-normalized the data to 1 at the effective wavelength of the V filter (540 nm), measured by a simple linear interpolation between $g$ and $r$ magnitudes, and scaled it by the corresponding AKARI albedo:
\begin{equation}\label{eq:8}
    S(\alpha) =p(\alpha) \big(S^r(\alpha)/S^{V}(\alpha)\big).
\end{equation}
In Fig. 1 of \cite{beck2021}, it is clear that when using ``absolute reflectance'', the spectra at different phase angles are not normalized. To include this, we used their supplementary information to compute the average change of reflectance for $\alpha\in[0,30]$ degrees ($S'=-0.00093$ units of bidirectional reflectance deg$^{-1}$) and applied in the scaling mentioned above via
\begin{equation}
    p(\alpha) = p_{\rm AKARI}+\alpha S'.
\end{equation}

Finally, we computed the spectral slope for different phase angles using Eq. \ref{eq:8} fitting a linear function and computed its change with $\alpha$, $S_{\alpha}$.
\begin{figure}
\centering
\includegraphics[width=\hsize]{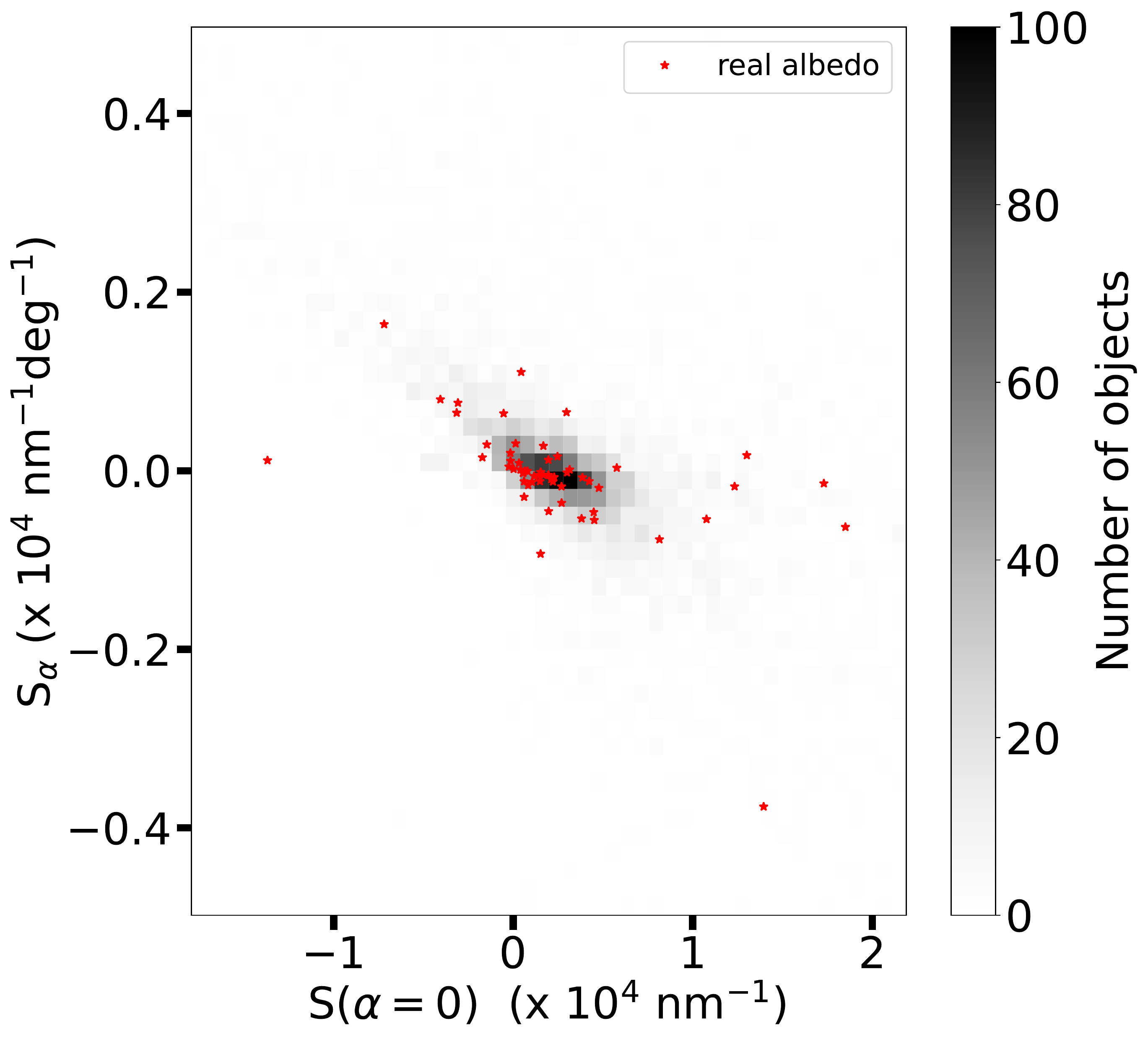}
\caption{Spectra slope at $\alpha=0$ deg vs. the rate of change of spectral slope with $\alpha$. Red asterisk show objects with measured $p_{AKARI}$ while the 2D histogram shows the complete dataset with random values of albedo. We do not show the error bars for clarity.}\label{fig:fig09}%
\end{figure}
We show the results as red asterisks in Fig. \ref{fig:fig09}, where we removed obvious outliers. The figure shows that objects with a blue slope (negative) at opposition ($S(\alpha=0)<0$ nm$^{-1}$) tend to get redder with increasing $\alpha$ ($S_{\alpha}>0$ nm$^{-1}$ deg$^{-1}$). Nevertheless, there are just a few data points. In order to increase the statistics, we use the complete sample of objects and compute $S^r(\alpha)$ for all possible objects. Because most objects do not have $p_{AKARI}$ measured, we will draw random values from the dataset and assume them as the ``real'' values. The rest of the process was identical as described before. We show the results as a two-dimensional histogram in the figure, showing that the correlation holds and is clearer. {\cite{sanchez2012} measured the effect of phase reddening on near-Earth asteroids and found that spectra of ordinary chondrite meteorites observed in laboratory change their slope very little below 30 deg, they report an initial slope in the range of $10^{-4}$ nm$^{-1}$ with a span in the order of $0.4\times10^{-4}$ nm$^{-1}$ deg$^{-1}$. Considering that our data include larger observational errors and different types of asteroids, the agreement in the order of magnitude with \cite{sanchez2012} is compelling, while the anti-correlation found in colors (and spectral slopes) seems real.}

Figure 9.12 of \cite{Grynko2008} shows numerical phase curves obtained for different density packing of a given material. The phase curve with a lower density is steeper than the higher packing density. How is this related to our work? Perhaps we see a variation in the packing density or the number of particles interacting with the light at different wavelengths. If shadow-hiding is the primary mechanism acting here, we are in the geometric optics domain and see particles bigger than the incident wavelength. Therefore, the effect we detect, the anti-correlation, is due to the size distribution of particles on the surfaces. 

\section{Conclusions}\label{sect:conclussions}
The main objective of this work was to discover if the phase curves of asteroids share the same anti-correlation found for TNOs using $H_V-H_R$ and $\Delta\beta$ \citep{ayala2018,alcan2019} if using the same photometric model (linear) and phase angle range. To improve the quality of our results, we applied the $P(H_{AC22})$ as priors. We used the same approach as in AC22 to obtain $>4000$ absolutes magnitudes for objects observed with $\alpha<7.5$ degrees. {We see that using previous results as priors in our new processing improves the results significantly, making the results obtained with a simple linear approach in low-$\alpha$ to resemble the results obtained with a full HG$_{12}^{*}$ model and complete range of $\alpha$ (Fig. \ref{fig:fig05}).}

We obtained strong correlations in consecutive pairs of colors, probably associated with the particle size distribution present on the surface of the objects. {The anti-correlations indicate that intrinsically redder objects become bluer with increasing phase angle, while the opposite happens for intrinsically bluer objects.} We checked that the correlation is not due to a normalization issue. When including taxonomy information, we see that the absolute color vs. $\Delta\beta$ space is covered slightly differently by the different main complexes, particularly the C and S-complex {although the anti-correlation holds for all taxa. The slightly different space covered by the C and S complexes resembles the parallel sequences seen in Fig. 17 of AC22, keeping in mind that their phase coefficients were $G^*_{12}$ and not $\beta$.}
To conclude, we suggest that the term phase-reddening should be changed to the more generic phase-coloring because some objects tend to become bluer with increasing phase angle.

\begin{acknowledgements}
We thank the input provided by the referee who improved this manuscript. 

AAC acknowledges support from the State Agency for Research of the Spanish MCIU through the ``Center of Excellence Severo Ochoa'' award to the Instituto de Astrofísica de Andalucía (SEV-2017-0709).

Funding for the creation and distribution of the SDSS Archive has been provided by the Alfred P. Sloan Foundation, the Participating Institutions, the National Aeronautics and Space Administration, the National Science Foundation, the U.S. Department of Energy, the Japanese Monbukagakusho, and the Max Planck Society. The SDSS Web site is http://www.sdss.org/.

The SDSS is managed by the Astrophysical Research Consortium (ARC) for the Participating Institutions. The Participating Institutions are The University of Chicago, Fermilab, the Institute for Advanced Study, the Japan Participation Group, The Johns Hopkins University, the Korean Scientist Group, Los Alamos National Laboratory, the Max-Planck-Institute for Astronomy (MPIA), the Max-Planck-Institute for Astrophysics (MPA), New Mexico State University, University of Pittsburgh, University of Portsmouth, Princeton University, the United States Naval Observatory, and the University of Washington.

This work is partially based on data from the SVO MOC Data Access Service at CAB (CSIC-INTA). This work used https://www.python.org/,
https://www.scipy.org/, and
Matplotlib \citep{hunte2007}. 

\end{acknowledgements}

% WARNING
%-------------------------------------------------------------------
% Please note that we have included the references to the file aa.dem in
% order to compile it, but we ask you to:
%
% - use BibTeX with the regular commands:
%   \bibliographystyle{aa} % style aa.bst
%   \bibliography{Yourfile} % your references Yourfile.bib
%
% - join the .bib files when you upload your source files
%-------------------------------------------------------------------
 \bibliographystyle{aa}
 \bibliography{hbeta}

\begin{thebibliography}{28}
\expandafter\ifx\csname natexlab\endcsname\relax\def\natexlab#1{#1}\fi

\bibitem[{{Al{\'\i}-Lagoa} {et~al.}(2018){Al{\'\i}-Lagoa}, {M{\"u}ller},
  {Usui}, \& {Hasegawa}}]{alilagoa2018}
{Al{\'\i}-Lagoa}, V., {M{\"u}ller}, T.~G., {Usui}, F., \& {Hasegawa}, S. 2018,
  \aap, 612, A85

\bibitem[{{Alvarez-Candal} {et~al.}(2019){Alvarez-Candal}, {Ayala-Loera},
  {Gil-Hutton}, {Ortiz}, {Santos-Sanz}, \& {Duffard}}]{alcan2019}
{Alvarez-Candal}, A., {Ayala-Loera}, C., {Gil-Hutton}, R., {et~al.} 2019,
  \mnras, 488, 3035

\bibitem[{{Alvarez-Candal} {et~al.}(2022){Alvarez-Candal}, {Benavidez}, {Campo
  Bagatin}, \& {Santana-Ros}}]{alcan2022}
{Alvarez-Candal}, A., {Benavidez}, P.~G., {Campo Bagatin}, A., \&
  {Santana-Ros}, T. 2022, \aap, 657, A80

\bibitem[{{Ayala-Loera} {et~al.}(2018){Ayala-Loera}, {Alvarez-Candal}, {Ortiz},
  {Duffard}, {Fern{\'a}ndez-Valenzuela}, {Santos-Sanz}, \&
  {Morales}}]{ayala2018}
{Ayala-Loera}, C., {Alvarez-Candal}, A., {Ortiz}, J.~L., {et~al.} 2018, \mnras,
  481, 1848

\bibitem[{Beck {et~al.}(2021)Beck, Schmitt, Potin, Pommerol, \&
  Brissaud}]{beck2021}
Beck, P., Schmitt, B., Potin, S., Pommerol, A., \& Brissaud, O. 2021, Icarus,
  354, 114066

\bibitem[{{Belskaya} {et~al.}(2006){Belskaya}, {Ortiz}, {Rousselot}, {Ivanova},
  {Borisov}, {Shevchenko}, \& {Peixinho}}]{belska2006}
{Belskaya}, I.~N., {Ortiz}, J.~L., {Rousselot}, P., {et~al.} 2006, \icarus,
  184, 277

\bibitem[{{Belskaya} \& {Shevchenko}(2000)}]{belskayashev2000}
{Belskaya}, I.~N. \& {Shevchenko}, V.~G. 2000, \icarus, 147, 94

\bibitem[{{Carry} {et~al.}(2016){Carry}, {Solano}, {Eggl}, \&
  {DeMeo}}]{carry2016Icar}
{Carry}, B., {Solano}, E., {Eggl}, S., \& {DeMeo}, F.~E. 2016, \icarus, 268,
  340

\bibitem[{{Colazo} {et~al.}(2022){Colazo}, {Alvarez-Candal}, \&
  {Duffard}}]{colazo2022}
{Colazo}, M., {Alvarez-Candal}, A., \& {Duffard}, R. 2022, arXiv e-prints,
  arXiv:2204.05075

\bibitem[{{Fern{\'a}ndez-Valenzuela} {et~al.}(2017){Fern{\'a}ndez-Valenzuela},
  {Ortiz}, {Duffard}, {Morales}, \& {Santos-Sanz}}]{estela2017Bienor}
{Fern{\'a}ndez-Valenzuela}, E., {Ortiz}, J.~L., {Duffard}, R., {Morales}, N.,
  \& {Santos-Sanz}, P. 2017, \mnras, 466, 4147

\bibitem[{{Gehrels}(1956)}]{Gehrels1956A}
{Gehrels}, T. 1956, \apj, 123, 331

\bibitem[{Grynko \& Shkuratov(2008)}]{Grynko2008}
Grynko, Y. \& Shkuratov, Y.~G. 2008, Light scattering from particulate surfaces
  in geometrical optics approximation, ed. A.~A. Kokhanovsky (Berlin,
  Heidelberg: Springer Berlin Heidelberg), 329--382

\bibitem[{{Hapke}(2002)}]{hapke2002}
{Hapke}, B. 2002, \icarus, 157, 523

\bibitem[{{Hapke}(1963)}]{hapke1963JGR}
{Hapke}, B.~W. 1963, \jgr, 68, 4571

\bibitem[{{Harris} \& {Lupishko}(1989)}]{harrislupishkp1989aste}
{Harris}, A.~W. \& {Lupishko}, D.~F. 1989, in Asteroids II, ed. R.~P. {Binzel},
  T.~{Gehrels}, \& M.~S. {Matthews}, 39--53

\bibitem[{{Hicks} {et~al.}(2005){Hicks}, {Simonelli}, \& {Buratti}}]{hicks2005}
{Hicks}, M.~D., {Simonelli}, D.~P., \& {Buratti}, B.~J. 2005, \icarus, 176, 492

\bibitem[{Hunter(2007)}]{hunte2007}
Hunter, J.~D. 2007, Computing In Science \& Engineering, 9, 90

\bibitem[{{Ivezi{\'c}} {et~al.}(2001){Ivezi{\'c}}, {Tabachnik}, {Rafikov},
  {Lupton}, {Quinn}, {Hammergren}, {Eyer}, {Chu}, {Armstrong}, {Fan},
  {Finlator}, {Geballe}, {Gunn}, {Hennessy}, {Knapp}, {Leggett}, {Munn},
  {Pier}, {Rockosi}, {Schneider}, {Strauss}, {Yanny}, {Brinkmann}, {Csabai},
  {Hindsley}, {Kent}, {Lamb}, {Margon}, {McKay}, {Smith}, {Waddel}, {York}, \&
  {SDSS Collaboration}}]{ivezic2001AJ}
{Ivezi{\'c}}, {\v{Z}}., {Tabachnik}, S., {Rafikov}, R., {et~al.} 2001, \aj,
  122, 2749

\bibitem[{{Jester} {et~al.}(2005){Jester}, {Schneider}, {Richards}, {Green},
  {Schmidt}, {Hall}, {Strauss}, {Vand en Berk}, {Stoughton}, {Gunn},
  {Brinkmann}, {Kent}, {Smith}, {Tucker}, \& {Yanny}}]{jester2005AJ}
{Jester}, S., {Schneider}, D.~P., {Richards}, G.~T., {et~al.} 2005, \aj, 130,
  873

\bibitem[{{Juri{\'c}} {et~al.}(2002){Juri{\'c}}, {Ivezi{\'c}}, {Lupton},
  {Quinn}, {Tabachnik}, {Fan}, {Gunn}, {Hennessy}, {Knapp}, {Munn}, {Pier},
  {Rockosi}, {Schneider}, {Brinkmann}, {Csabai}, \& {Fukugita}}]{juric2002AJ}
{Juri{\'c}}, M., {Ivezi{\'c}}, {\v{Z}}., {Lupton}, R.~H., {et~al.} 2002, \aj,
  124, 1776

\bibitem[{{Lumme} \& {Bowell}(1981)}]{lummebowell1981AJ}
{Lumme}, K. \& {Bowell}, E. 1981, \aj, 86, 1705

\bibitem[{{Mahlke} {et~al.}(2021){Mahlke}, {Carry}, \& {Denneau}}]{mahlke2021}
{Mahlke}, M., {Carry}, B., \& {Denneau}, L. 2021, \icarus, 354, 114094

\bibitem[{{Muinonen}(1989)}]{muino1989OE}
{Muinonen}, K. 1989, \ao, 28, 3044

\bibitem[{{Muinonen} {et~al.}(2010){Muinonen}, {Belskaya}, {Cellino},
  {Delb{\`o}}, {Levasseur-Regourd}, {Penttil{\"a}}, \&
  {Tedesco}}]{muinonen2010HG1G2}
{Muinonen}, K., {Belskaya}, I.~N., {Cellino}, A., {et~al.} 2010, \icarus, 209,
  542

\bibitem[{{Penttil{\"a}} {et~al.}(2016){Penttil{\"a}}, {Shevchenko}, {Wilkman},
  \& {Muinonen}}]{penti2016HG}
{Penttil{\"a}}, A., {Shevchenko}, V.~G., {Wilkman}, O., \& {Muinonen}, K. 2016,
  \planss, 123, 117

\bibitem[{{Rabinowitz} {et~al.}(2007){Rabinowitz}, {Schaefer}, \&
  {Tourtellotte}}]{rabi2007AJ}
{Rabinowitz}, D.~L., {Schaefer}, B.~E., \& {Tourtellotte}, S.~W. 2007, \aj,
  133, 26

\bibitem[{{Sanchez} {et~al.}(2012){Sanchez}, {Reddy}, {Nathues}, {Cloutis},
  {Mann}, \& {Hiesinger}}]{sanchez2012}
{Sanchez}, J.~A., {Reddy}, V., {Nathues}, A., {et~al.} 2012, \icarus, 220, 36

\bibitem[{{Verbiscer} {et~al.}(2022){Verbiscer}, {Helfenstein}, {Porter},
  {Benecchi}, {Kavelaars}, {Lauer}, {Peng}, {Protopapa}, {Spencer}, {Stern},
  {Weaver}, {Buie}, {Buratti}, {Olkin}, {Parker}, {Singer}, {Young}, \& {New
  Horizons Science Team}}]{verbi2022}
{Verbiscer}, A.~J., {Helfenstein}, P., {Porter}, S.~B., {et~al.} 2022, The
  Planetary Science Journal, 3, 95

\end{thebibliography}

 \begin{appendix}
 \section{The linear approximation of a non-linear region}\label{appA}
 
The phase curve of small bodies (up to $\alpha\approx30$ degrees) can be roughly described as a linear behavior for $\alpha>10$ degrees, while for smaller angles, a non-linear behavior could appear especially in moderate to high albedo objects \citep{belskayashev2000}. This non-linear behavior was named Opposition Effect \citep[OE,][]{Gehrels1956A}. Therefore, if the phase curve is non-linear at low-$\alpha$, can we describe it adequately with a simple linear model?
 
To answer this question, we proceeded as follows: We assumed an object with $H=17$ (footnote \footnote{The actual absolute magnitude of our model is irrelevant as we only care for the differences between models.}) and used the HG$_{12}^{*}$ model to create three different phase curves using G$_{12}^{*}\in\{0.1,0.5,0.9\}$, keeping in mind that lower values of G$_{12}^{*}$ imply larger non-linear behaviors. From the modeled phase curves we extracted randomly $n$ pairs $(\alpha_i,m_i)$, $i$ from 1 to $n$, with $n=3,6,9,$ or 12. We used the same selection criterion described in the main text: at least 3 data. In this case, we did not include the effect of observational errors because they may blur the results further than desired.
 
\begin{figure}
\centering
  \includegraphics[width=4cm]{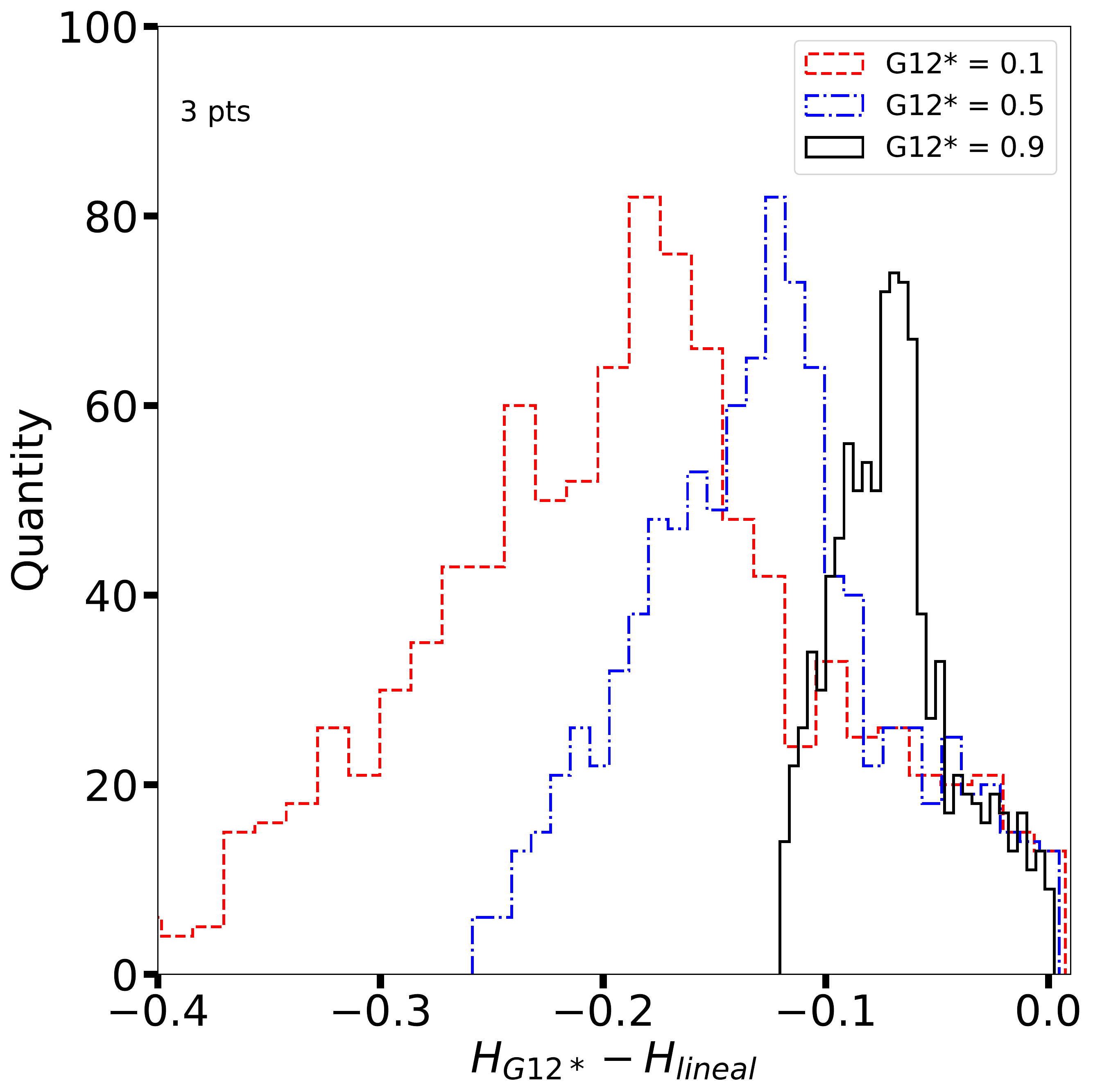}
  \includegraphics[width=4cm]{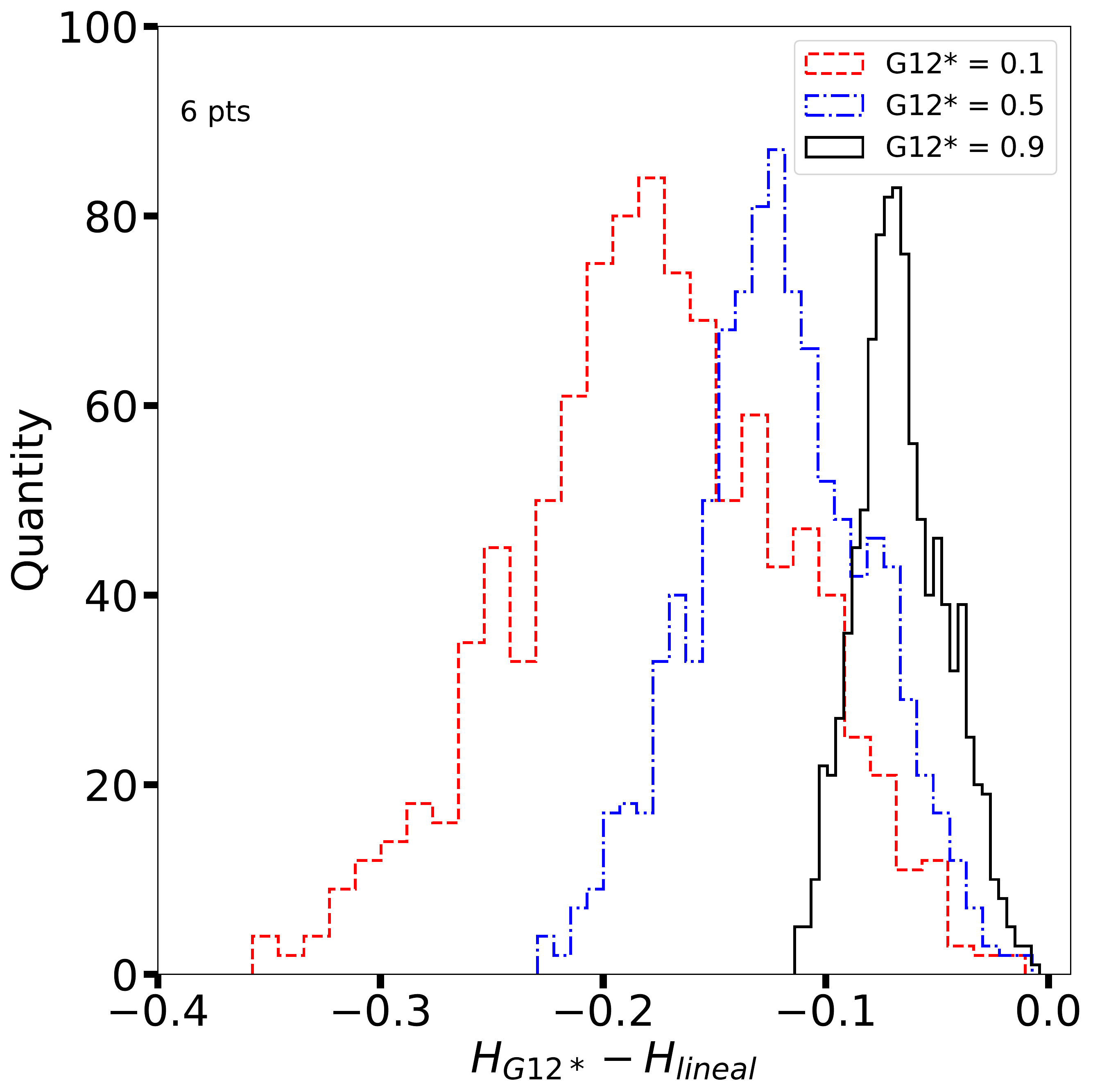}
  \includegraphics[width=4cm]{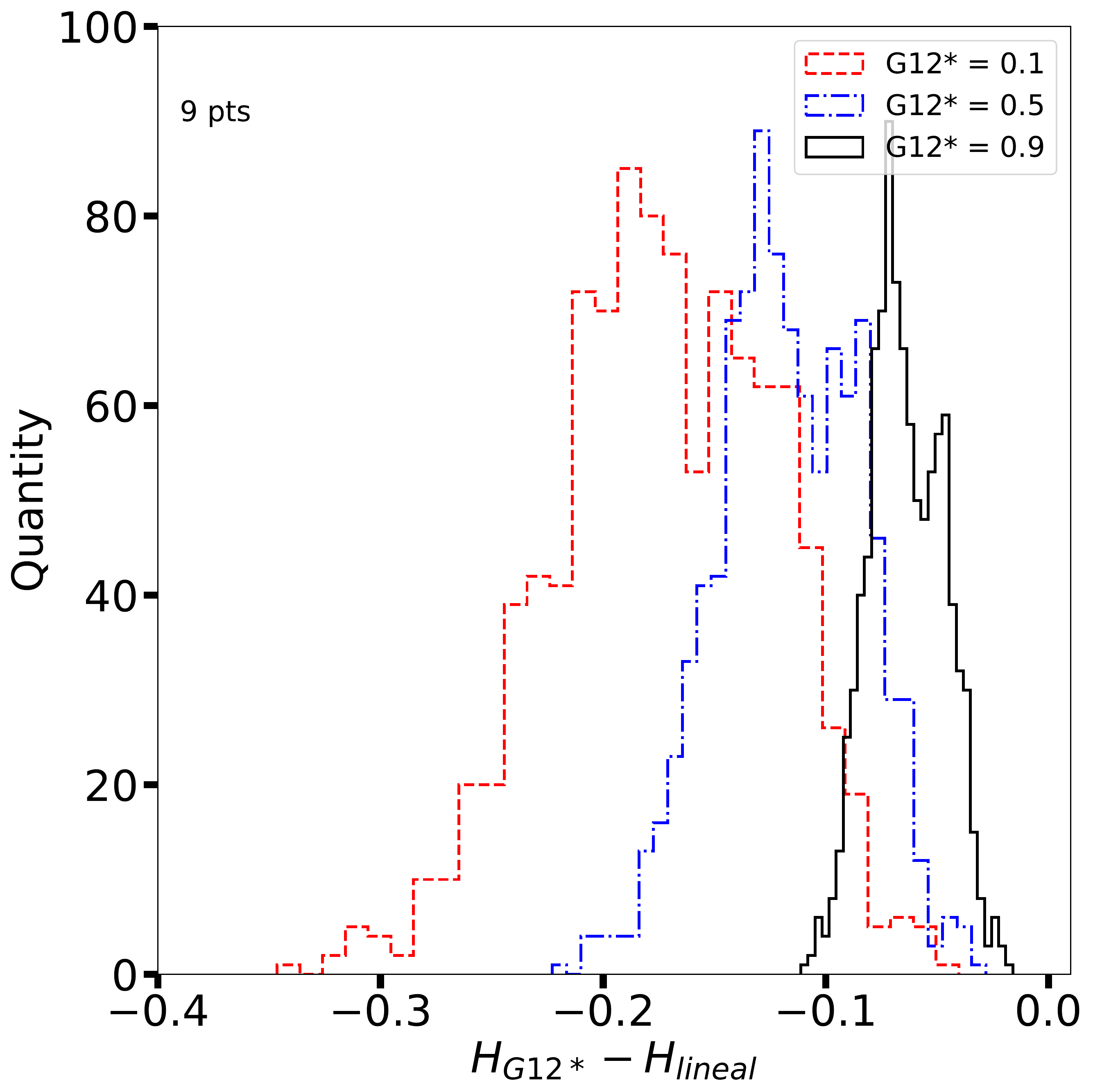}
  \includegraphics[width=4cm]{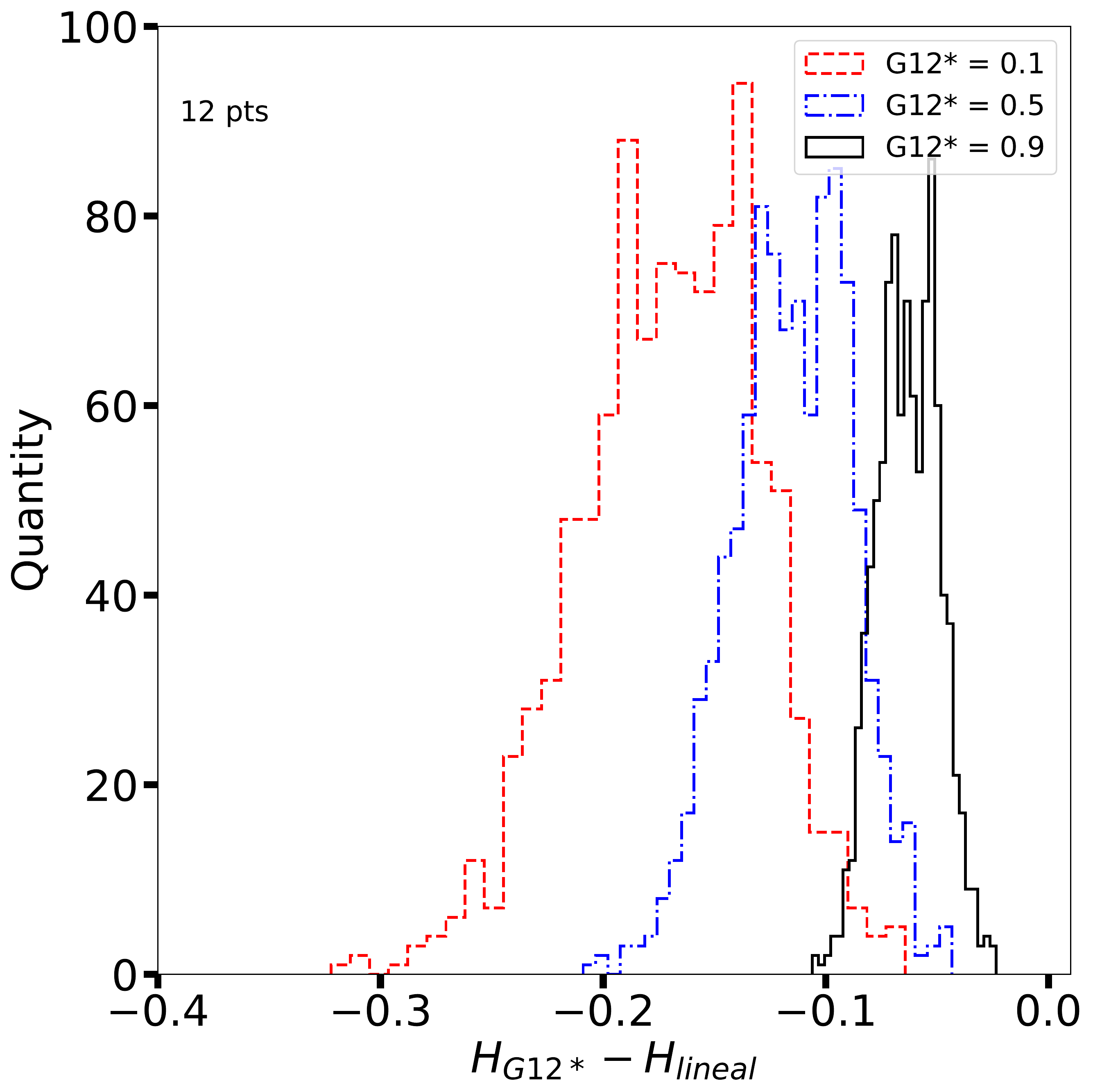}
\caption{$H_{G12*}-H_{lineal}$ for a modeled asteroid. Each panel is labeled (top left) with the number $n$ of pairs, while the histograms are labeled according to the value of G$_{12}^{*}$ used to draw the pairs (see text for details).}\label{fig:a01}%
\end{figure}

 Using the data $(\alpha_i,m_i)$ we computed linear fits to obtain $H_{linear}$, and then computed the difference $17-H_{linear}$. For each value of G$_{12}^{*}$ and $n$, we extracted 1000 different samples to create histograms and to avoid spurious results due to the random nature of our selection of samples. We show the results in Fig. \ref{fig:a01}, where we labeled the four panels with the number of pairs used and each histogram with its respective value of phase coefficient.
 
From the figure, we summarize: As expected, the stronger non-linear behavior shows larger discrepancies when using a linear model, with a maximum at about -0.4 mag, but with many results better than -0.2 mag. Curiously, with low $n$, the difference may decrease to about zero or be slightly positive; this does not mean that the linear model describes well the non-linear behavior, but that with a sparse coverage, the estimated absolute magnitude does not differ much from the one using more complex photometric models. With large $n$, this is not the case (as seen in the bottom panels of the figure), although the median differences for a given G$_{12}^{*}$ seems to be roughly independent of $n$ (Table \ref{table:ap1})
 \begin{table}
\caption{$H_{G12*}-H_{linear}$}
\label{table:ap1}
\centering
\begin{tabular}{c c | c c c |  c c c}
\hline\hline
& &Ast& & & TNOs& & \\
\hline
$n$ & G$_{12}^{*}$ & min & med & max & min & med & max \\
\hline
3& 0.1&-0.41&  -0.19&  0.01&-0.14& -0.07& 0.03\\ 
 & 0.5&-0.26&  -0.13&  0.00&-0.10& -0.05& 0.02\\
 & 0.9&-0.12&  -0.07&  0.00&-0.06& -0.03& 0.01\\
 \hline
6& 0.1&-0.36&  -0.18& -0.01&-0.13& -0.05& 0.01\\
 & 0.5&-0.23&  -0.12& -0.01&-0.09& -0.04& 0.01\\
 & 0.9&-0.11&  -0.07&  0.00&-0.05& -0.02& 0.01\\
 \hline
9& 0.1&-0.35&  -0.17& -0.04&-0.13& -0.05& 0.00\\
 & 0.5&-0.22&  -0.12& -0.03&-0.09& -0.04& 0.00\\
 & 0.9&-0.11&  -0.07& -0.02&-0.05& -0.02& 0.00\\
 \hline
2& 0.1&-0.32&  -0.17& -0.06&-0.12& -0.05& 0.00\\
 & 0.5&-0.21&  -0.11& -0.04&-0.09& -0.04& 0.00\\
 & 0.9&-0.11&  -0.06& -0.02&-0.05& -0.02& 0.00\\

\hline
\end{tabular}
%\tablefoot{}
\end{table}
In conclusion, the linear model is not adequate to study OE {\it per se}, but it provides absolute magnitudes in reasonable-to-good agreement with the HG$_{12}^{*}$ model.

\begin{figure}
\centering
 \includegraphics[width=4cm]{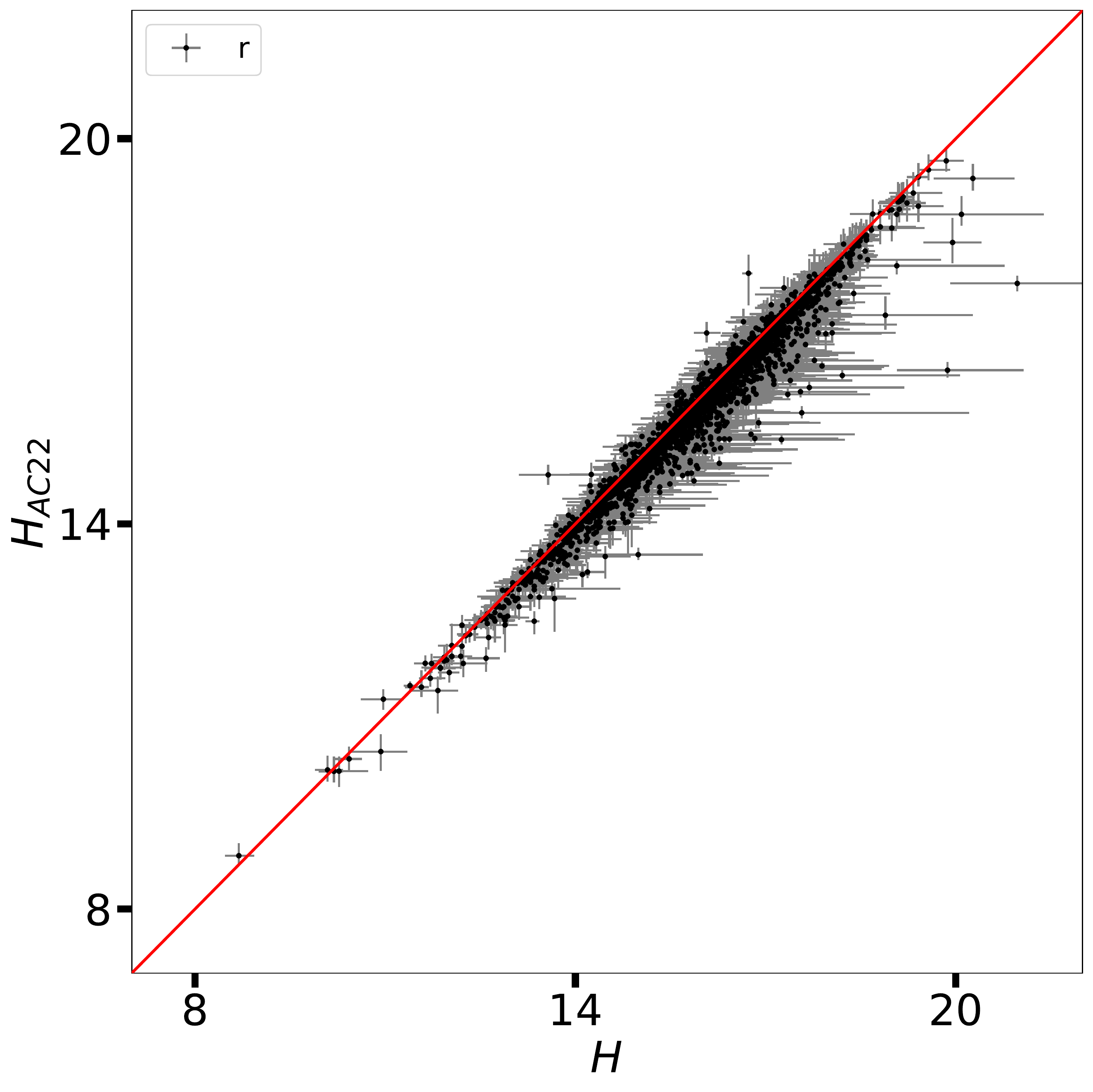}
 \includegraphics[width=4cm]{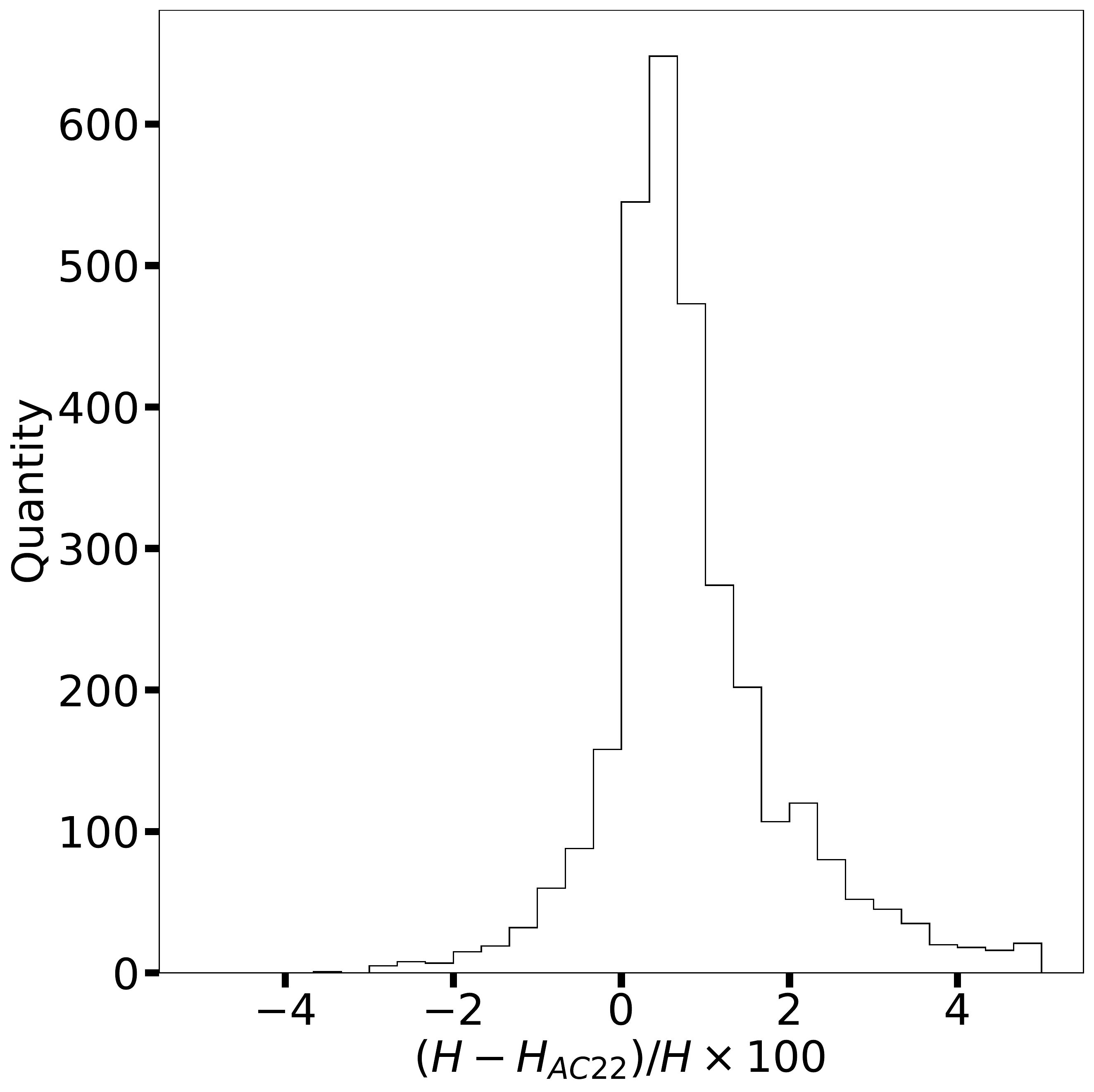}
\caption{Comparison between linear and HG$^*_{12}$ models. Left panel: $H_r$ obtained by us using the linear model (labeled $H$) against the magnitude obtained in AC22. The red line indicates the $1:1$ relation. Right panel: Distribution of the percentage change between the obtained $H$; we cut the x-axis for clarity.
}\label{fig:fig5}%
\end{figure}

To assess the impact of a simple linear model applied to our data, we compare our $Hs$ with AC22's (Fig. \ref{fig:fig5}).
As already seen in Fig. \ref{fig:a01}, the linear approach underestimates the absolute magnitudes, which is also the case for the actual data with a median difference of 0.10. {Generally, the difference is less than $2-2.5$ \%}. The difference between the actual and modeled data is that the actual data account for observational errors and possible rotational states, as well as priors have been applied, which made the final magnitudes closer to these obtained with the HG$_{12}^{*}$ model. The results are similar in all filters. We have checked that the median difference is roughly independent of the minimum $\alpha$ and the coverage of the phase curve. {Considering all this, we confirm that assuming a linear behavior, although it underestimates $H$, it fits the majority of objects satisfactorily enough.}

\begin{figure}
\centering
 \includegraphics[width=4cm]{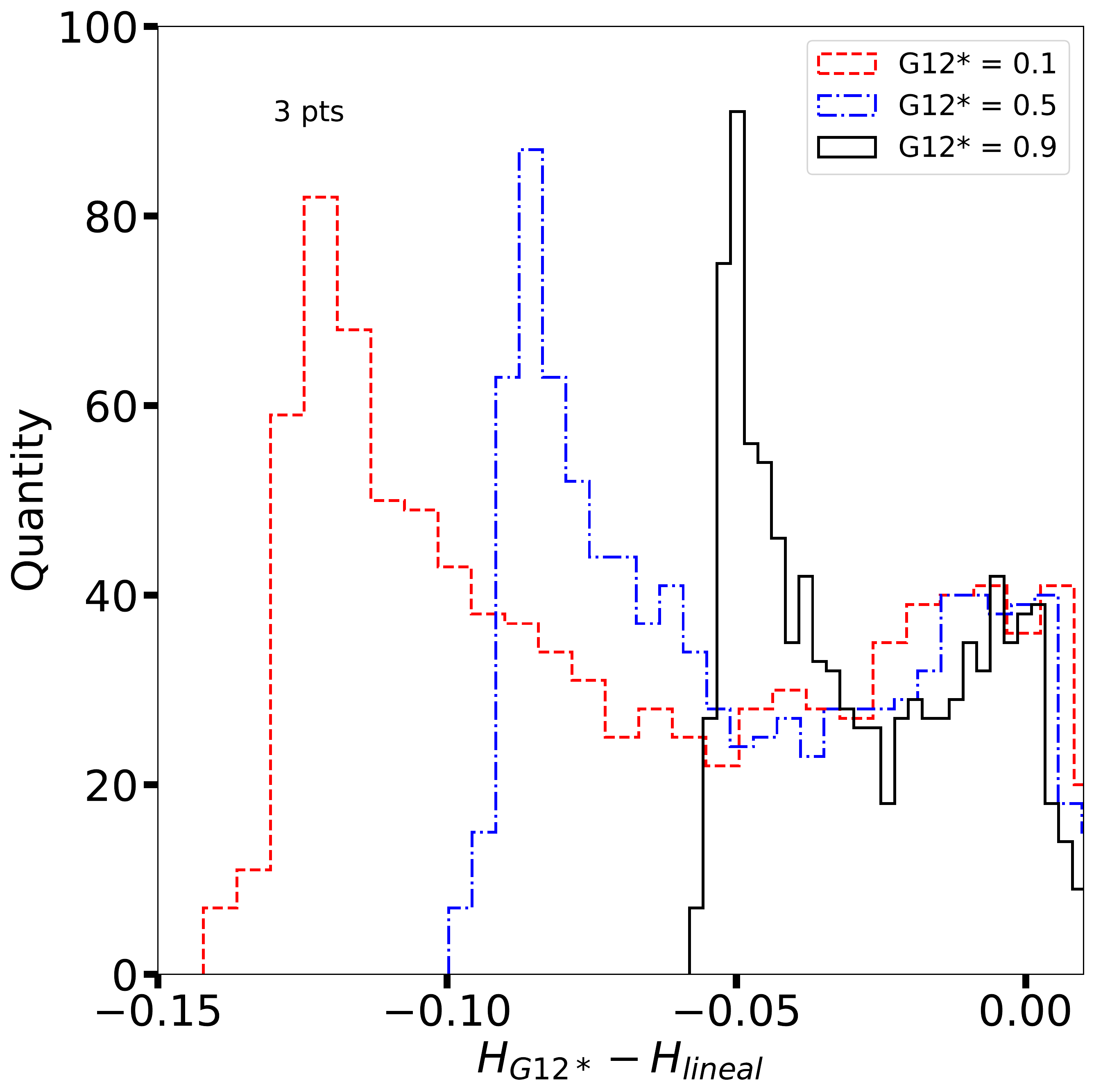}
 \includegraphics[width=4cm]{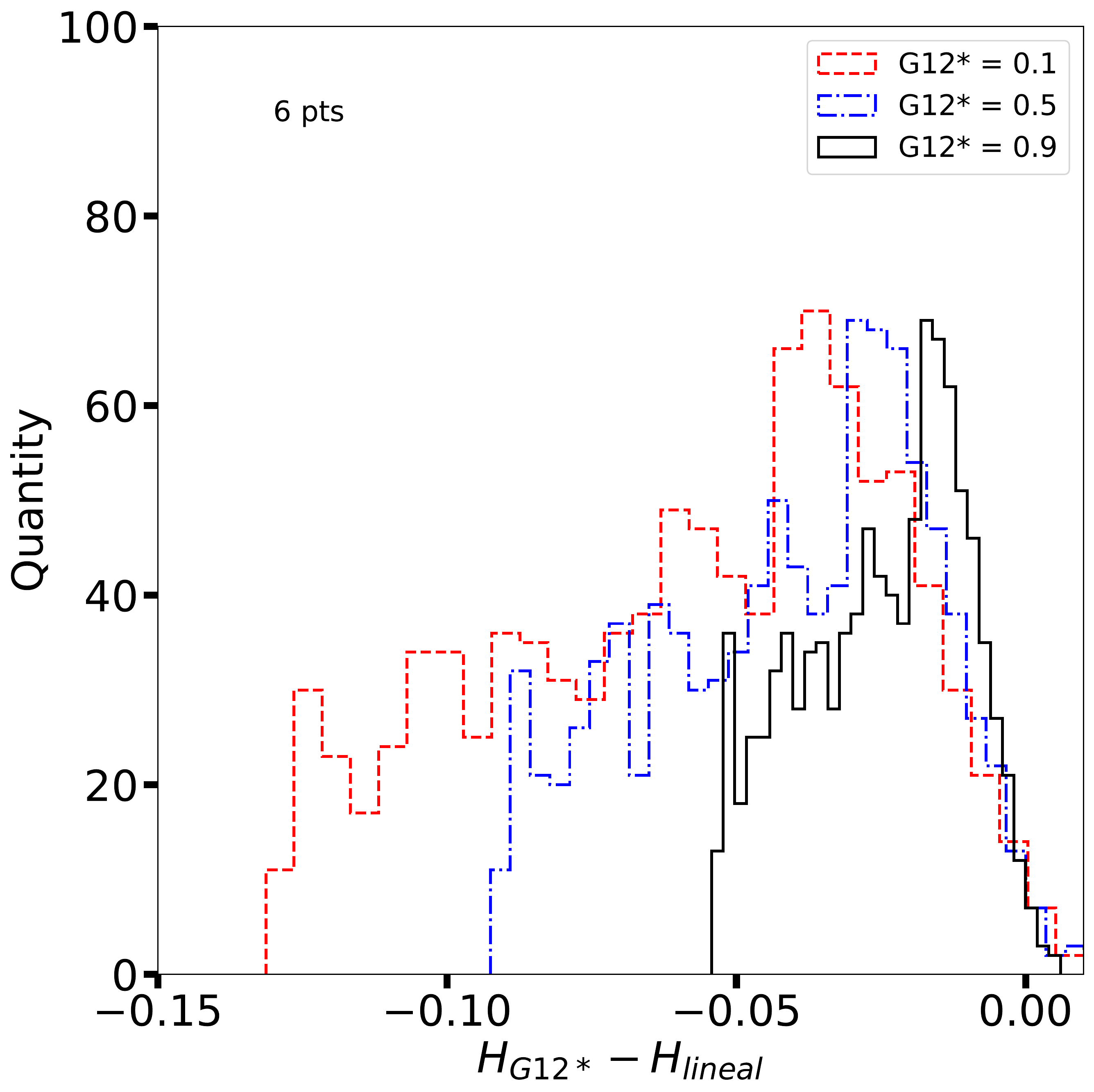}
 \includegraphics[width=4cm]{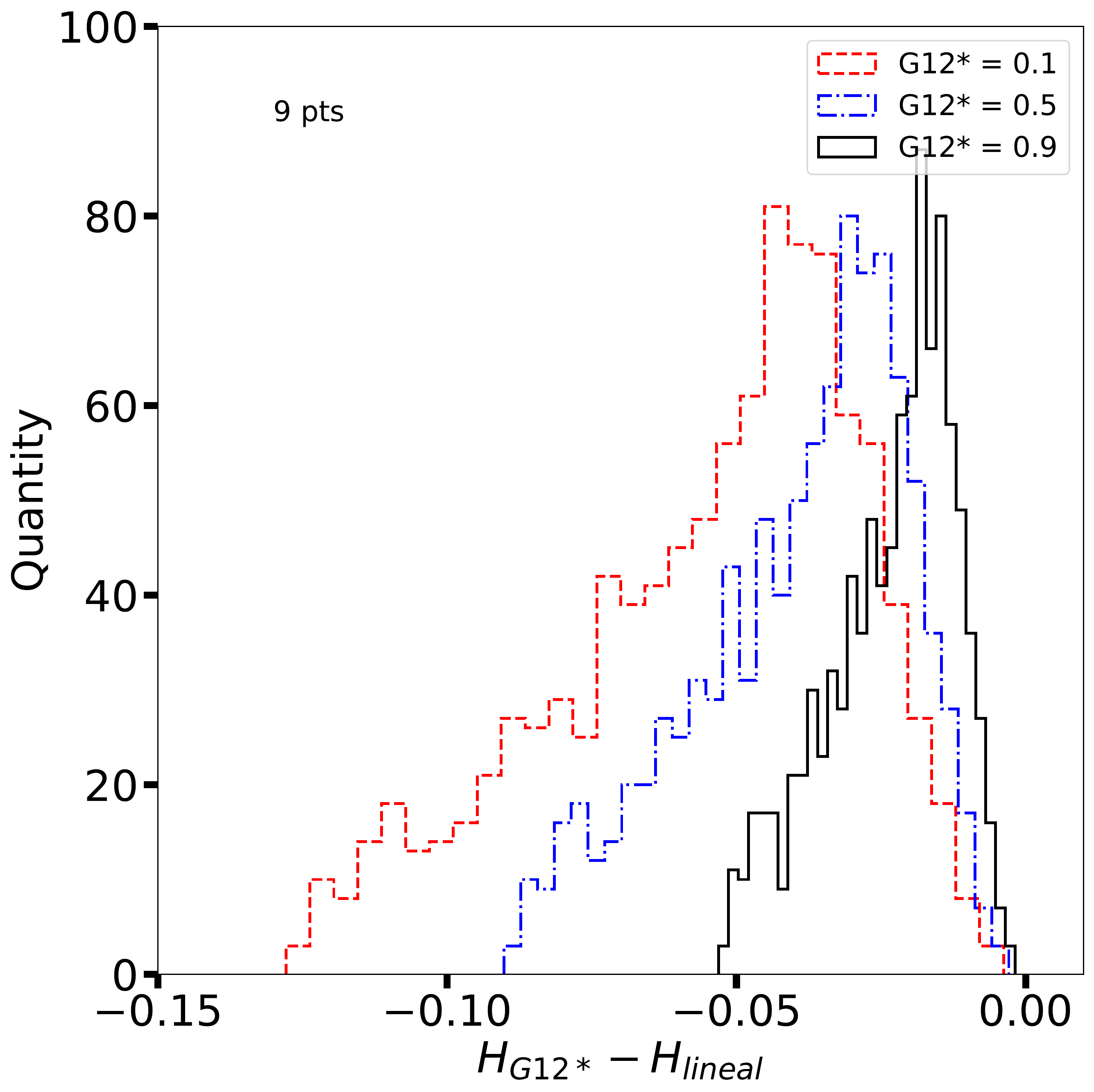}
 \includegraphics[width=4cm]{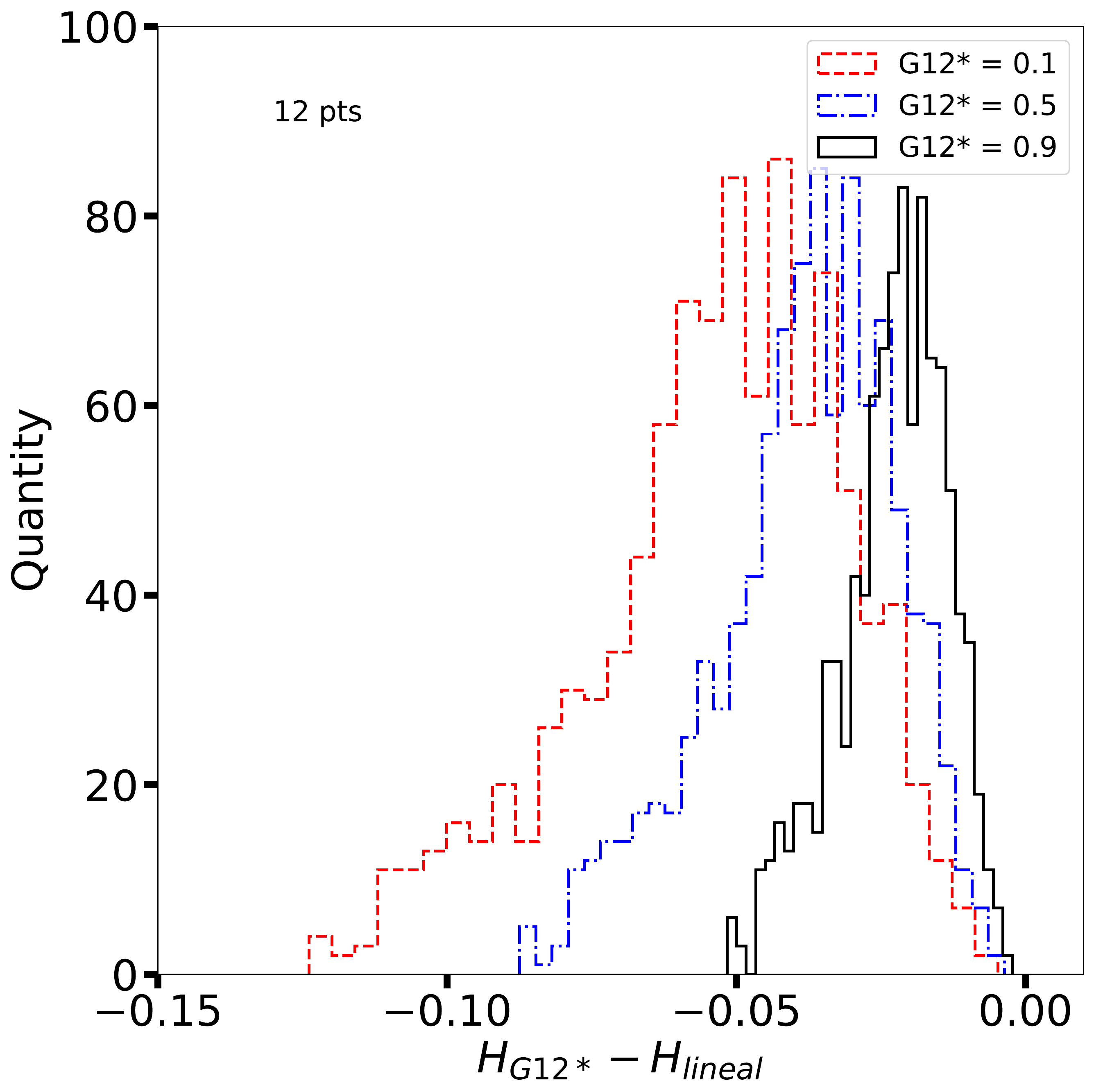}
\caption{$H_{G12*}-H_{lineal}$ for a modeled trans-Neptunian object. Each panel is labeled (top left) with the number $n$ of pairs, while the histograms are labeled according to the value of G$_{12}^{*}$ used to draw the pairs (see text for details).}\label{fig:a02}%
\end{figure}

\smallskip{}
As a plus, we computed the same effect but considering trans-Neptunian objects, i.e., maximum $\alpha=2$ degrees and no other constraint. The processing was the same as above, and we show the results in Fig. \ref{fig:a02}. 
In this case, the linear approach works much better than in asteroids due to the restricted phase angle range, with a maximum difference of about -0.15 mag. The shape of the distribution for $n=3$ is different from the other $n$ because, with such a small phase angle coverage, three values of $m_i$ close together may account for a large difference in the linear parameters than better-spaced values. The linear approximation works well in most cases, which justifies its use in ground-based data of TNOs. 

Nevertheless, we must keep in mind that a few TNOs/centaurs show non-linear behavior at a low-phase angle: Bienor \citep{rabi2007AJ}, Varuna \citep{hicks2005,belska2006}, Pluto, Charon, and Triton \citep{verbi2022}. Except for Charon, all other objects have (or may have) associated phenomena that affect their photometric behavior: Varuna has a large amplitude of the rotational light curve, Bienor has an odd photometric behavior \citep{estela2017Bienor}, and Pluto and Triton have atmospheres.

 \end{appendix}

%\begin{thebibliography}{}
%\end{thebibliography}
\end{document}